\documentclass[
p
reprint,
pre,
showpacs,
showkeys,
longbibliography
]{revtex4-1}
\usepackage[displaymath, mathlines]{lineno}
\usepackage{amsmath, amssymb, graphicx,amsfonts,bm
}
\newcommand{\ER}[0]{Erd\H{o}s-R\'{e}nyi }
\newcommand{\dstr}{\bm{\pi}}
\newcommand{\fig}[2]{Fig.~\ref{#1}#2}
\usepackage{subfigure}

\newcommand{\comment}[1]{}

\begin{document}

\title{Spatial networks evolving to reduce length}%

\author{Chris Varghese}
\email{varghese@phy.duke.edu}
\affiliation{Department of Physics, Duke University, Durham, North Carolina, USA}

\author{Rick Durrett}
\email{rtd@math.duke.edu}
\affiliation{Department of Mathematics, Duke University, Durham, North Carolina, USA}


\date{\today}
\begin{abstract}
Motivated by results of Henry, Pralat and Zhang (PNAS 108.21 (2011): 8605-8610), we propose a general scheme for evolving spatial networks in order to reduce their total edge lengths. We study the properties of the 
equilbria of two networks from this class, which interpolate between three well studied objects: the Erd\H{o}s-R\'{e}nyi random graph, the random geometric graph, and the minimum spanning tree. 
The first of our two evolutions can be used as a model for a social network where individuals have fixed opinions about a number of issues and adjust their ties to be connected to people with similar views. 
The second evolution which preserves the connectivity of the network has potential applications in the design of transportation networks and other distribution systems.
\end{abstract}

\maketitle

\section{Introduction}
The availability of real world network data spurred enormous interest in the
study of complex networks starting in the late 1990s \cite{newman03}. Numerous
models have been proposed for the formation of many observed 
technological, social, information and biological networks. Many of these
network models were purely topological, i.e., the location of the vertices of
the network were irrelevant. 
However, it is clear that most real world networks have a spatial element to
them. 
Examples include transportation networks
\cite{Xie2007,Aldous2008,Aldous2008a,Tero2010,Louf2013}, distribution networks
\cite{Gastner2006}, 
some social networks \cite{Frasco2013} and the neural network in
the brain \cite{Bullmore2009, Bullmore2012a, Bassett2011b}. See
\cite{Barthelemy2011} for an extensive review. The effects of space on the
topology can be significant.  
For example, in a social network, individuals are likely to have more friends
closer to their spatial locations than farther away. 

Many of the models of spatial network that have been proposed are essentially
static. Well studied models of this nature include 
the random geometric graph, the Waxman model of the internet \cite{Waxman1988},
and the Watts-Strogatz model \cite{watts98} that generates small-world networks.
Barnett, Paolo and Bullock \cite{Barnett2007} performed an 
extensive study of networks where the probability that vertex pairs are
connected depend on their spatial separation.
Frasco et al.~\cite{Frasco2013} studied a model for the formation of social
networks where the topology was decided first 
and vertices were then sequentially placed in a square depending on the topology
and distance to already placed vertices.

However, most real networks are not static but rather evolve in an attempt to
improve their efficiency. For example, the networks in the brain are constantly rewired 
during the human life span for cognitive development and improving other brain functions \cite{sporns10,boccaletti06}.

\subsection{A general evolution scheme}
We consider the equilibrium of evolving undirected spatial networks $G_t =
(V,E_t)$, where $V = V_{nD}$ is a
set of $n$ points uniformly distributed in a $D$-dimensional space 
$\mathcal{V}_{nD}$ of volume $n$, so that the mean density of points is unity.
Although any metric can in principle be used, we will stick to the familiar
Euclidean metric for defining distances.
With applications to transportation and distribution networks in mind, the
boundaries of the space are not periodic. We are primarily interested in the
``thermodynamic'' limit $n \to \infty$ and 
dimension $D=2$; so $\mathcal{V}_{n2}$ is, say, a square of side $\sqrt{n}$.
The network evolves only through the rewiring of edges, so the number of edges
$|E_t|$ at time $t$, and consequently, the mean degree $\mu = 2|E_t|/n$ are
constant.

Consider a spatial network as defined above that is required to satisfy some
topological constraint $\mathcal{T}$ 
(we only consider the constraint that the network is connected; however, other
examples include: the network is planar, 
the degree of the vertices is bounded, etc.), and evolve with the aim of
lowering its total length. 
Assume that the edges rewire independently of each other and according to the
following Metropolis-Hastings dynamics \cite{Metropolis1953}:

\begin{itemize}
\item Edges attempt to rewire at a rate proportional to some
power $\delta \geq 0$ of their length. So if $\delta >0$, larger edges have a
higher
tendency to attempt to rewire. 
\item If a proposed rewiring of an edge of length $ \ell$ to an edge of length
$\ell'$ leads to a network that satisfies constraint $\mathcal{T}$, 
then it is accepted with a probability $\min(1 , f(\ell)/f(\ell') )$, where
$f(\cdot)$ is a non-decreasing function. In other words, 
a shorter edge is definitely 
accepted while the chance of a longer edge getting accepted decreases with length.   
\end{itemize}

With the above evolution scheme, it is easy to find the distribution of the
networks at equilibrium. 
Consider the set $\mathcal{G}$ of possible networks. Let
$G,F \in \mathcal{G}$ and suppose that $F$ is formed by rewiring edge $\{x, y\}$
in $G$ to
$\{x, z\}$. Without loss of generality assume $| x - y | > | x - z | $. The
transition rates of going from $G$ to $F$ and from $F$ to $G$ respectively in
one step are
\begin{align}
\Lambda(G \rightarrow F) &= | x-y|^\delta \,\frac{1}{n-1 - d(x)}\,, \quad
\textrm{ and } \nonumber\\ 
\Lambda(F\rightarrow G ) &= | x-z|^\delta \, \frac{1}{n-1 - d(x)}
\frac{f(|x-z|)}{f(|x-y|)}\, , 
\end{align}
where $d(x)$ is the degree of vertex $x$.
We seek an equilibrium distribution $\dstr$ that satisfies detailed balance
\begin{align}
\dstr(G) \, \Lambda(G \rightarrow F) = \dstr(F) \,\Lambda(F\rightarrow G)\,.
\end{align}
This holds if $\dstr(G)$ is proportional to 
\begin{align}
\prod_{\{x,y\}\in E} \frac{ 1 }{| x-y|^\delta f(| x-y|)} = 
\exp\left[ - \sum_{\{x,y\}\in E} \log\left(| x-y|^\delta f(| x-y|)\right)\right]
\,.\label{mouse}
\end{align}
Note that all the transition rates and probabilities above are conditional on
the vertex set $V_{nD}$. One may interpret the distribution 
\eqref{mouse} as follows: the cost of an edge is an increasing function of its
length $\ell$, specifically, $\log [\ell^\delta f(\ell)]$; 
the cost of a network is the sum of the cost of its edges; 
the equilibrium networks have a distribution that is exponential in their cost.

The main motivation for our work is the model of segregation in a social network
by Henry, Pra\l at and Zhang (HPZ) \cite{Henry2011}, 
which corresponds to the case $\delta >0, \, f(x) = \textrm{constant}$\comment{in the general evolution scheme}. 
They defined their model in discrete time with
a parameter $p$ that controls the rate of convergence to equilibrium.  
Motivated by HPZ, Magura et al.~\cite{Magura2013} studied a continuous time
model with $\delta=1$ and $f(x) = x^{\alpha-1}$.

\subsection{Our model}
In order to have short edges, we choose $f(x) = e^{\beta x}$, where $\beta$ is a
non-negative parameter. For simplicity, we set $\delta=0$.
In other words, edges make independent rewire attempts at a constant rate 1, and
longer edges are accepted with a probability that decays exponentially with the
difference in the lengths. 
Thus, in our model, at each evolution step: 
an edge $\{x,y\} \in E$ is chosen at random and one of its vertices, say, $x$ is
designated as its pivot; 
the vertex $x$ chooses another vertex $z$ outside its neighborhood;  
if the network created by rewiring the edge $\{x,y\}$ to $\{x, z\}$ satisfies
constraint $\mathcal{T}$, 
then the move is accepted with probability $\min\left[1 ,
e^{-\beta(|x-z|-|x-y|)} \right]$.

Substituting $\delta =0$ and $f(x) = e^{\beta x}$ in \eqref{mouse}, we find our
equilibrium network to be in the set
$\mathcal{G}(V_{nD},\mu,\mathcal{T})$ of spatial
networks with vertex set $V_{nD}$ and $n \mu/2$ edges that satisfy constraint
$\mathcal{T}$, and with a probability measure 

\begin{equation}
 \dstr(G) = \frac{1}{Z_{\beta \mu}}  e^{ -\beta H(G )}\,, \label{eqdist}
\end{equation}
where $ H(G) = \sum_{\{x,y\} \in E(G)}| x-y|$ is the total length of the
network, and $Z_{\beta \mu}=  \sum_{G\in \mathcal{G}} e^{ -\beta H(G )}$ is a
normalization constant. 
Thus, in going from the general evolution scheme to our model, we have made the
definition of the cost of an edge more specific, i.e., the cost is proportional
to its length, with the cost per unit length being $\beta$.
With $n\to\infty$, the four parameters $D$, $\beta$, $\mu$ and $\mathcal{T}$
specify the equilibrium network of our Evolving Spatial Network model which we
abbreviate as ESNM. The first two parameters $D$ and $\beta$ 
control the spatial effects, while the latter two -- $\mu$ and $\mathcal{T}$
regulate the topology of the network.

\section{The unconstrained network}
In the simplest version of our model, the network is not required to satisfy any
constraint. With this simplification, as we see below, the model is closely
related to a percolation process and hence is 
amenable to some analytical calculations.

\subsection{A Fermion gas picture and connection with percolation}
\label{EquilibriumNetwork}
In the unconstrained network, the distribution \eqref{eqdist} of the equilibrium
network leads us to an
alternative view of the model. 
If we treat the $\binom{n}{2}$ 
possible vertex pairs $\{x,y\}$ as the single particle energy levels $|x-y|$ in
a Fermionic system, and the edges of the network to correspond to the occupied
energy levels, then we have a non-interacting Fermionic system 
(constraints on the network would mean interacting Fermions). The parameter
$\beta$ may then be viewed as the inverse temperature, $H(G)$ as the Hamiltonian
of the system, and $Z_{\beta \mu}$ as the canonical partition function. 
However, having a fixed number of edges (canonical ensemble description) is inconvenient for computations, 
so we will use a grand canonical ensemble description which
is equivalent to that of the canonical ensemble when the number
particles is large. Given $V_{nD}$, the grand canonical partition function is
\begin{align}
 \Xi(\beta,\kappa) = \sum_{G \in \mathcal{G}(V_{nD})} \kappa^{|E(G)|} e^{-\beta
H(G)}\,,
\end{align}
where $\kappa$ is the fugacity and $\mathcal{G}(V_{nD})$ is the set of simple
graphs with vertex set $V_{nD}$.

Another way to look at the equilibrium network, which is equivalent to the
grand canonical description above, is to view it as the result of a percolation
process on $V_{nD}$. 
For this, consider the set $\mathcal{G}(V_{nD})$ of graphs as before, but now
the edges assigned independently between all vertex pairs $\{x,y\}$ with
probability $g(|x-y|)$. 
Barnett, Paolo and Bullock \cite{Barnett2007} studied such percolation networks
for arbitrary functions $g(\cdot)$ and called them Spatially Embedded Random
Networks.
The distribution $\dstr'$ of the percolation network is 
\begin{align}
\dstr'(G) &= \prod_{\{x,y\} \in E} g(| x-y|) \prod_{\{x,y\} \in \binom{V}{2}
\setminus E} (1-g(| x-y|)) \nonumber\\ 
&= \left[ \prod_{\{x,y\} \in \binom{V}{2}} (1-g(| x-y|))  \right]\prod_{\{x,y\}
\in E} \frac{g(| x-y|)}{1-g(| x-y|) } \,,
\end{align}
where $\binom{V}{2}$ is the set of vertex pairs. This distribution can be made
similar to $\dstr$ if we let
\begin{align}
 \frac{g(\varepsilon)}{1-g(\varepsilon) }  =  \kappa e^{-\beta \varepsilon}
\quad\hbox{which means} \quad g(\varepsilon) = \frac{1}{1 +\kappa^{-1}e^{\beta
\varepsilon}}\,.
\end{align}
As shown in \cite{Magura2013}, 
the properties of the percolation model are closely related to those of the
random graph model if we choose $\kappa$ such that the expected mean degree in
the percolation network is equal to the mean degree $\mu$ of our model. 
The only difference is that while the number of edges is fixed in the ESNM, it
is random in the percolation version. 

\subsection{Properties of the percolation network}
\begin{figure}[h]
\centering
\subfigure[\comment{\ Distribution of edge lengths with $\mu=5$, $D=2$ for various values of $\beta$.}]{\includegraphics
[width=.47\textwidth]
{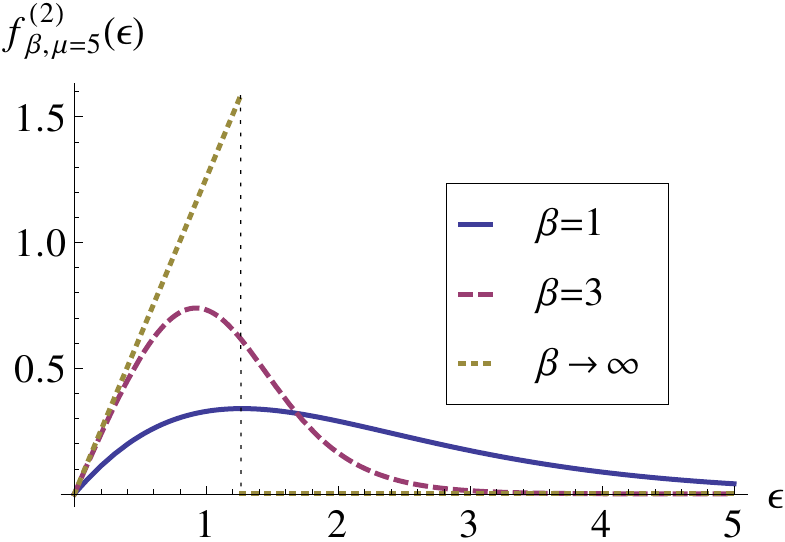} \label{EdgeLengthDistribution}}\,
\subfigure[\comment{\ Mean edge length $\xi$ a function of $\mu$ when $D=2$ for various values of $\beta$.}]{\includegraphics
[width=.47\textwidth]
{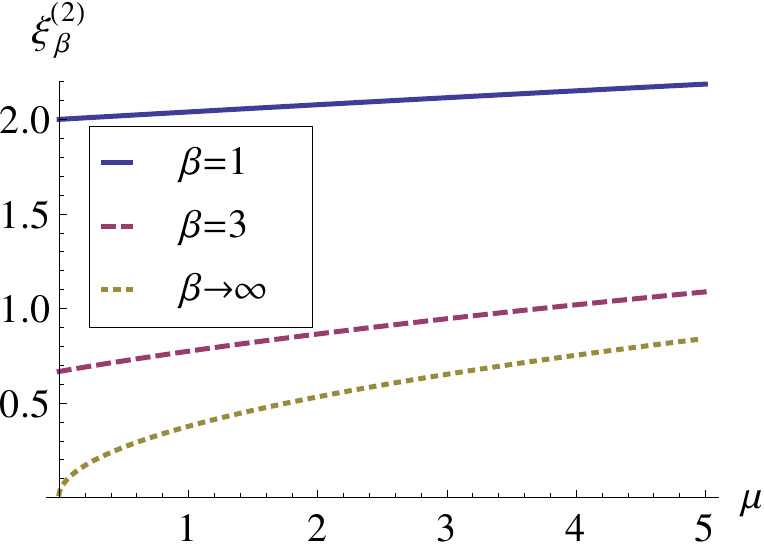} \label{EnergyDensity}}
\caption{Results for $\beta = 1$ (solid line), 3 (dashed line), and $\infty$ (dotted line) in dimension $D=2$; (a) gives the distribution of edge lengths when $\mu = 5$, (b) gives the mean edge length as a function of $\mu$.}
\label{EdgeLengths}
\end{figure}

In Appendix \ref{chlorine}, we find $\mathbb{E} \log \Xi$ from which the following two quantities can be directly calculated. 
First, the expected number of edges in the grand canonical
ensemble is 
\begin{align}
 \mathbb{E} |E(G)|  = \mathbb{E} [\mathbb{E} [ |E(G)| \,|V_{nD}]  ]
= \mathbb{E} \frac{\partial \log \Xi}{\partial \log \kappa } 
&=\frac{\partial \mathbb{E} \log \Xi}{\partial \log \kappa } \nonumber\\
&= \frac{n}{2}\frac{S_{D-1}}{\beta^D} \,\Gamma(D) \, \left[-\mathrm{Li}_D
(-\kappa )
\right]\,,
\end{align}
where $S_{D-1} = D \pi^{\frac{D}{2}}/\Gamma(1 + \frac{D}{2})$ is the area of the
unit $(D-1)$-sphere, and $\mathrm{Li}_s(z) = \sum_{k=1}^\infty z^k/k^s $ is the
Polylogarithm function. In order that the grand canonical ensemble description
be equivalent to the ESNM, we need to set $\mathbb{E} |E(G)|$ equal to the
number of edges $n\mu/2$ in the ESNM. This means 
\begin{align}
 \mu = \frac{S_{D-1}}{\beta^D} \,\Gamma(D) \, \left[-\mathrm{Li}_D (-\kappa )
\right] \,.\label{KappaDef}
\end{align}
In the rest of the section we will treat $\kappa = \kappa_{\beta\mu}^{(D)}$ to
be implicitly defined through \eqref{KappaDef}, 
and $g(\cdot) = g_{\beta\mu}^{(D)}(\cdot)$. 
\comment{For example in one dimension, $\kappa_{\beta\mu}^{(1)} = e^{\beta \mu/2}-1$ and
$g_{\beta\mu}^{(1)}(\varepsilon) = [1+ e^{\beta \varepsilon}/(e^{\beta
\mu/2}-1)]^{-1} $.}

Second, the expected value of the network Hamiltonian, i.e., the expected total
length of the network is
\begin{align}
 \mathbb{E} H(G) = -  \frac{  \partial \mathbb{E} \log \Xi}{\partial \beta} =
\frac{n }{2} \frac{S_{D-1}\Gamma(D+1)}{\beta^{D+1}} \left[-\mathrm{Li}_{D+1}
\left(-\kappa_{\beta\mu}^{(D)} \right)
\right]\,.
\end{align}
The mean edge length $\xi = \mathbb{E} [ H(G)/|E(G)| ]$. When $n\to \infty$,
both $ H(G)/n$ and $|E(G)|/n$ will converge to their respective limits, so that
 \begin{align}
  \xi = \xi_{\beta \mu}^{(D)} \to  \frac{ \mathbb{E} H(G) }{ \mathbb{E} |E(G)| }
= \frac{S_{D-1}\Gamma(D+1)}{\mu \beta^{D+1}} \left[-\mathrm{Li}_{D+1}
\left(-\kappa_{\beta\mu}^{(D)} \right)\right]\,.
\end{align}

To find the distribution of vertex degrees and edge lengths it is more convenient to use  the percolation picture. 
Since the neighbors of a vertex are assigned independently of each other with a probability that depends on distance, the degree distribution is Poisson (for a proof see Appendix \ref{sodium}).
Next we consider the distribution of the lengths of the edges in the network. We
want to find the probability 
\begin{align}
 \mathbb{P}(|x-y|=\varepsilon | \{x,y\} \in E) &= \frac{ \mathbb{P}(\{x,y\} \in
E
||x-y|=\varepsilon ) \mathbb{P}(|x-y|=\varepsilon) }{\mathbb{P}(\{x,y\} \in
E)}\nonumber\\
&= \frac{g_{\beta\mu}^{(D)}(\varepsilon ) \mathbb{P}(|x-y|=\varepsilon)}{\int
g_{\beta\mu}^{(D)}(\varepsilon' ) \mathbb{P}(|x-y|=\varepsilon')} \,.\label{cat}
\end{align}
As $n\to\infty$, $\mathbb{P}(|x-y|=\varepsilon) \to S_{D-1}
\varepsilon^{D-1}\mathrm{d}\varepsilon/n$. Substituting in \eqref{cat}, we get
the probability density function of the distribution of 
edge lengths to be (see Fig. \ref{EdgeLengthDistribution})
\begin{align} \label{crimea}
 f_{\beta\mu}^{(D)}(\varepsilon) = S_{D-1} \frac{g_{\beta\mu}^{(D)}(\varepsilon
)  \varepsilon^{D-1}}{\mu} \,.
\end{align}

\begin{figure}[h]
\centering
\subfigure[$\beta=0$]{\includegraphics
[width=.45\textwidth]
{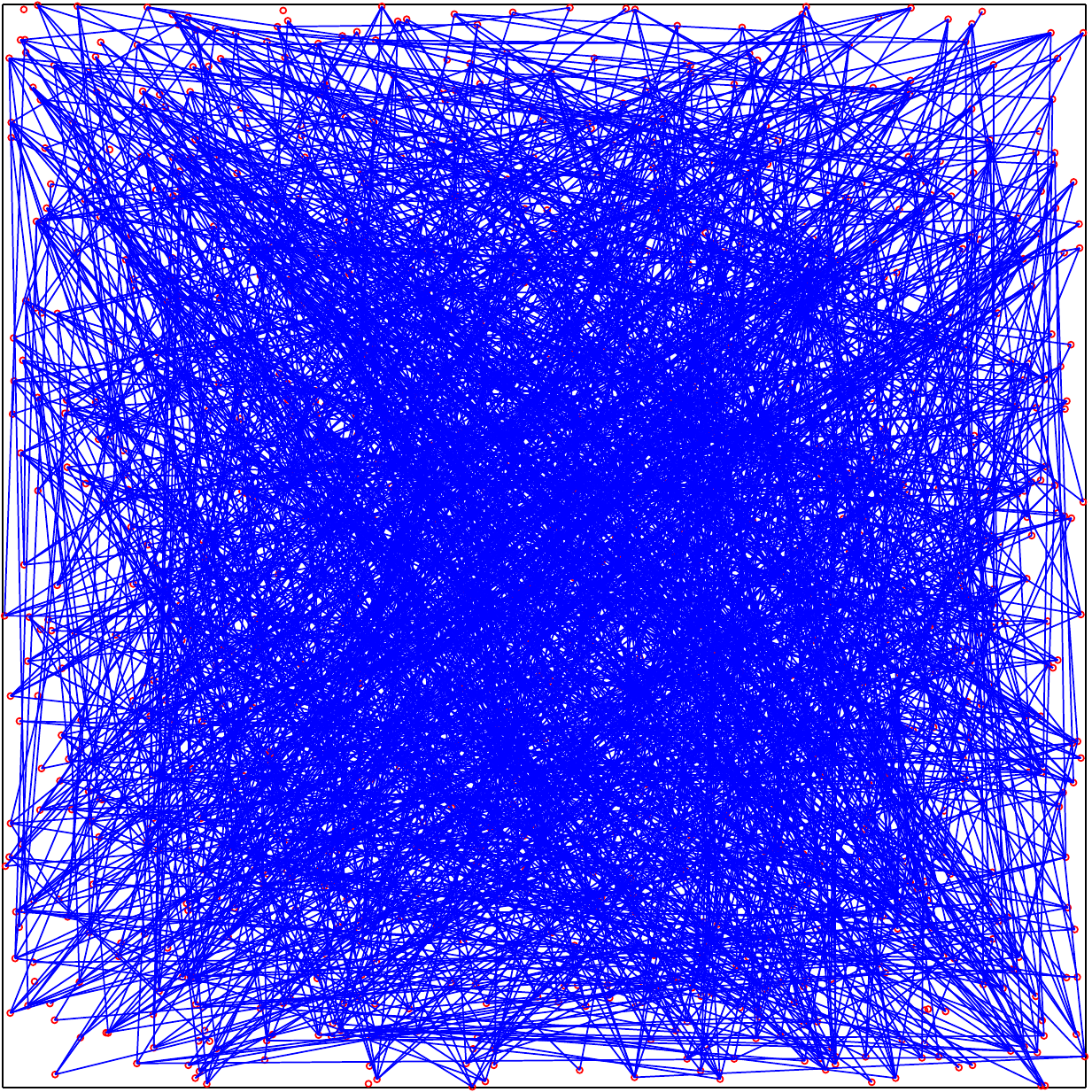} \label{INMsn3mu4D2b0k0sOUTpicture}
}\,
\subfigure[$\beta \to \infty$]{\includegraphics
[width=.45\textwidth]
{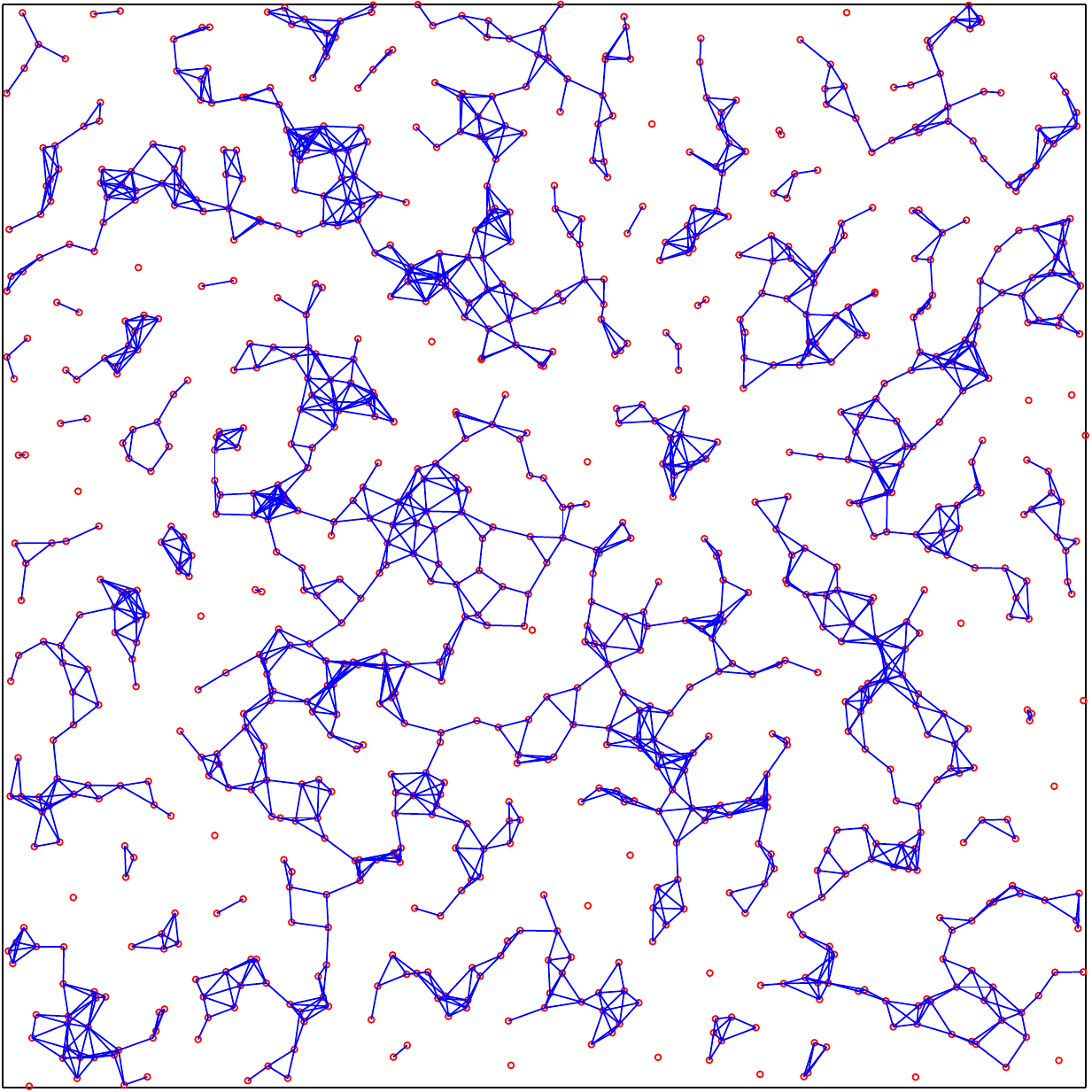} \label{INMsn3mu4D2binfk0sOUTpicture}}
\caption{Pictures of the (a) \ER graph, and (b) the random geometeric graph, on a square with $n=1000$ vertices.}
\label{INMsn3mu4D2b0binfk0sOUTpicture}
\end{figure}

The two extreme values of $\beta$ deserve special consideration. With $\beta
=0$, the spatial location of the vertices have no effect on the
evolution of the network and so the equilibrium network is the
Erd\H{o}s-R\'{e}nyi graph (a random graph drawn uniformly from the set of
networks with a given number of vertices and edges) of which the degree
distribution, 
the formation of the giant component, clustering coefficient, etc. are well
known \cite{newman10}. As $n\to \infty$, the mean edge length grows as the mean
vertex pair distance which is $\mathcal{O}(n^{1/D})$. Note that this is
consistent with the fact that $\lim_{\beta\to 0} \xi_{\beta \mu}^{(D)} =
\infty$ for all $D$, $\mu>0$.

In the limit of large $\beta $, the rewiring algorithm becomes a greedy
algorithm that always chooses the shorter edge. 
The equilibrium network will then be a random geometric graph (RGG)
\cite{Dall2002,Penrose2003} where vertices are connected to all their spatial
neighbors up to a  
distance $\varepsilon_{0}$, equal to the $|E|$-th smallest distance between
vertices. Alternatively, $\varepsilon_0$ is the Fermi energy 
\footnote{Recall that the Fermi energy of a Fermionic system is the highest
occupied single particle energy level at zero temperature.}
of the system. This means 
\begin{align}
 \binom{n}{2} \mathbb{P}(|x-y|<\varepsilon_{0}) &= |E| \,.
\end{align}
So as $n\to \infty$, we have
\begin{align}
\binom{n}{2}   \int_0^{\varepsilon_{0}} \frac{1}{n}S_{D-1}
\varepsilon^{D-1}\mathrm{d}\varepsilon = n\frac{\mu}{2} \quad \hbox{which means}
\quad \Omega_D \varepsilon_{0}^{D} = \mu  \,,
\end{align}
where $\Omega_D = \pi^{\frac{D}{2}}/\Gamma(1+ D/2)$
is the volume
of a unit 
$D$- ball. The mean edge length is 
\begin{align}
  \xi_{\beta\to\infty,\mu}^{(D)} = \int_{0}^{\varepsilon_{0}} \varepsilon \,
\frac{ S_{D-1} \varepsilon^{D-1} \mathrm{d}\varepsilon }{\Omega_D
\varepsilon_{0}^D} =
\frac{S_{D-1}}{(D+1) \Omega_D} \varepsilon_{0} =
\frac{D}{(D+1)\Omega_D^{\frac{1}{D}}} \mu^{1/D} \,.\label{horse}
\end{align}
 
The clustering coefficient $C$ is defined as the probability that two vertices 
that are connected to a common third vertex are also connected to each other, i.e., for three randomly chosen vertices $x$, $y$ and $z$, 
$
 C = \mathbb{P}(\{x,z\}\in E|  \{x,y\},\{y,z\}\in E)
$. Spatial networks should be expected to have high clustering, since two spatial neighbors of a vertex are also spatial neighbors of
each other.
For the percolation network, the clustering coefficient is 
\begin{align}
 C = \frac{\int \int g(|x|) \, g(|y|)\, g(|x-y|) \, \mathrm{d}x \, \mathrm{d}y }{\int \int g(|x|) \, g(|y|)\, \mathrm{d}x \, \mathrm{d}y}\,. \label{titanic}
\end{align}
For general values of $\beta$, it is difficult to evaluate \eqref{titanic}. However, in the $\beta\to\infty$ limit, $C$ can be calculated as given in \cite{Dall2002} and found to be (see Appendix \ref{potassium})
\begin{align}
 C_{\beta\to\infty,\mu}^{(D)} &= \frac{2D^2}{\sqrt{\pi}} \frac{\Gamma(D/2)}{\Gamma((D+1)/2)} \int_0^1  \int_0^{\arccos(t/2)} \sin^D \tau \,\mathrm{d}\tau \, t^{D-1}  \,\mathrm{d}t \,.\label{cow}
\end{align}
Notice that $C_{\beta\to\infty,\mu}^{(D)}$ is independent of the mean degree $\mu$.

\begin{figure}[h]
\centering
\subfigure[]{\includegraphics
[width=.45\textwidth]
{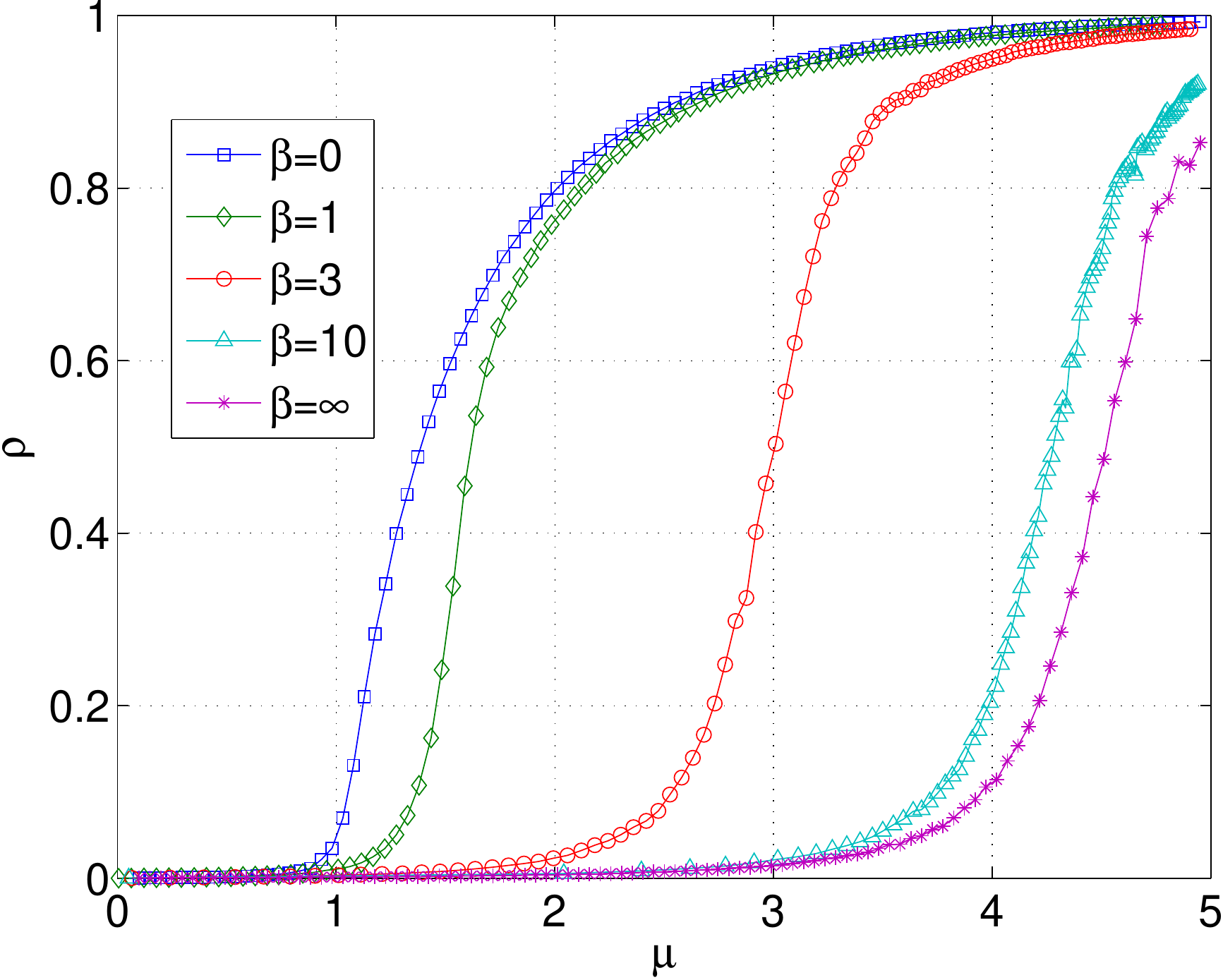}
\label{INn4mu0D2b0b1b3b10binfsOUTc}}\,
\subfigure[]{\includegraphics
[width=.45\textwidth]
{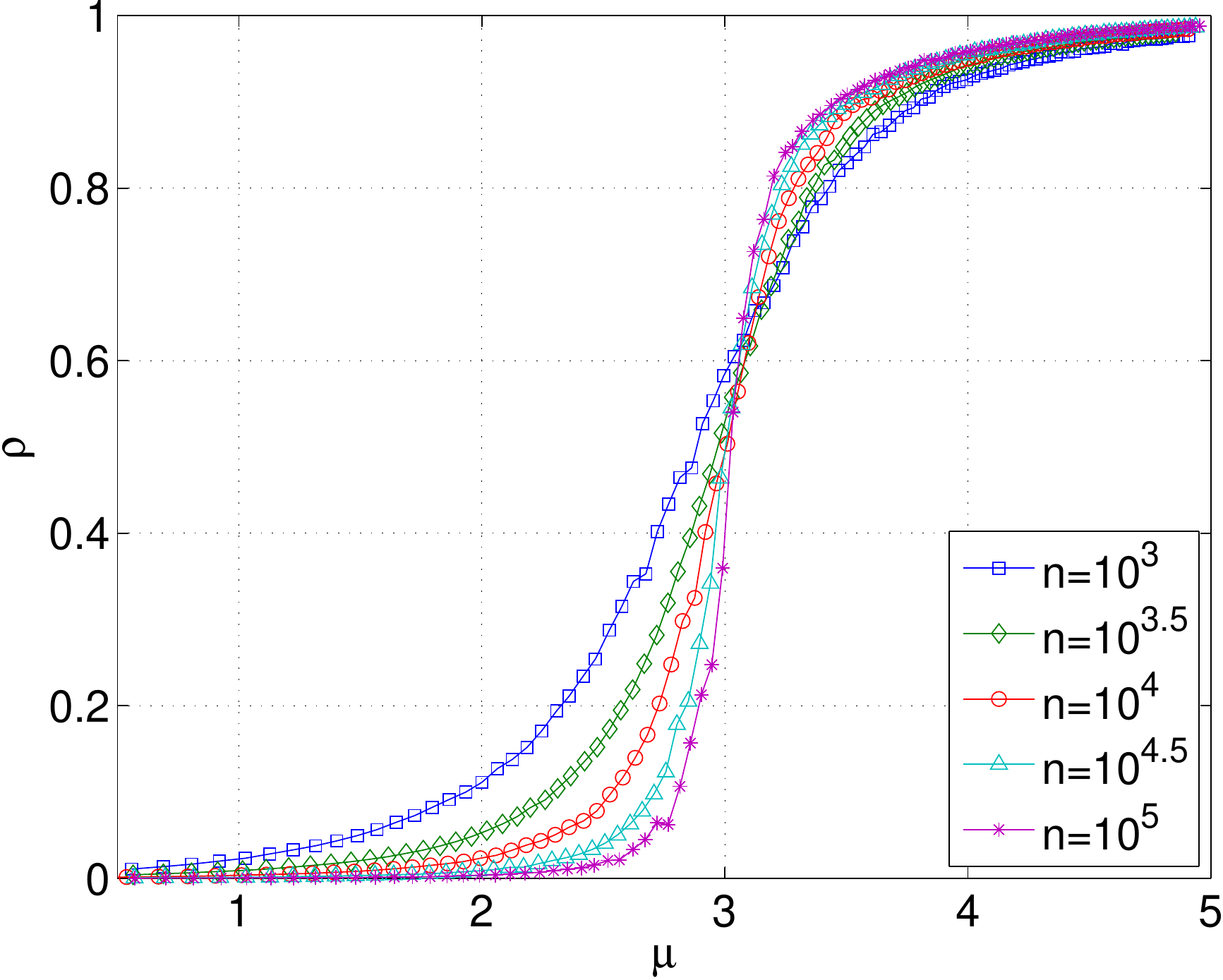}
\label{INn3n3_5n4n4_5n5mu0D2b3sOUTc} }\\
\subfigure[\comment{$\rho$ as a function of $\beta$ for various values of
$\mu$}]{\includegraphics
[width=.45\textwidth]
{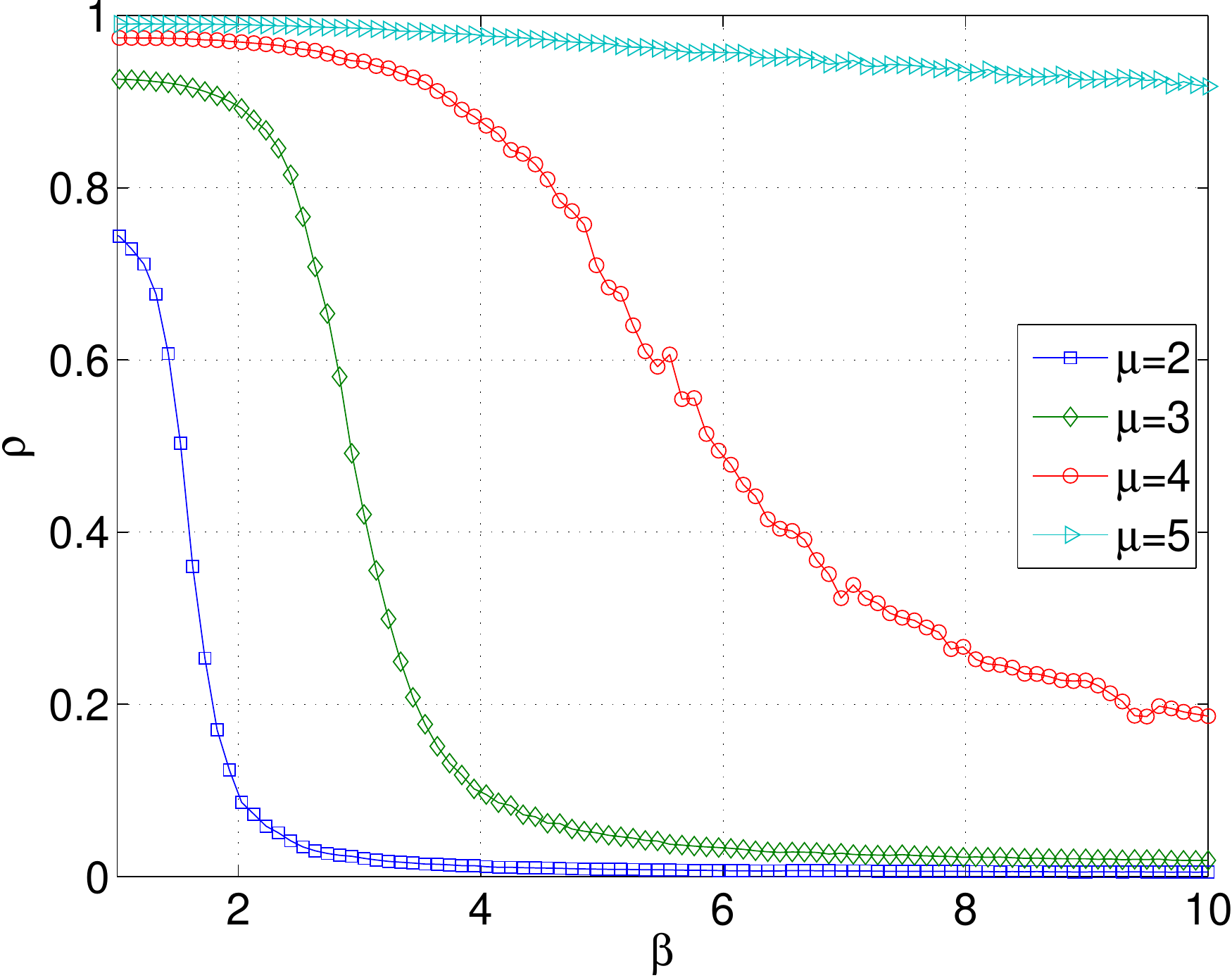}
\label{INMsn4mu2mu3mu4mu5D2sOUTc} }
\caption{Fraction of vertices $\rho$ in the largest component (a) as a function
of $\mu$ for various values of $\beta$ and (c) as a function of $\beta$ for
various values of $\mu$. (b) shows the finite size scaling for $\beta=3$; notice
that all the curves seem to cross at one point.}
\label{GiantComponent}
\end{figure}
It does not seem possible to analytically compute the size of the giant
component in the percolation network. 
Because of this, we simulate the percolation process. In all the results that
follow, the dimension 
$D=2$, and, unless otherwise stated, the network size $n =10^4$.

For fixed $\beta$, when $\mu$ is varied, the equilibrium network undergoes a
percolation transition (in the $n\to \infty$ limit), indicated by the fraction
$\rho$ of vertices 
in the giant component. 
For $\beta =0$, we know that the critical mean degree $\mu_{{*}}^{(\beta=0)} =1$
for formation of the giant component (see \fig{INn4mu0D2b0b1b3b10binfsOUTc}{}). 
Increasing $\beta$ makes the formation of the largest component difficult, as
long connections are not favored. 
However, there is an upper bound on $\mu_{{*}}$ achieved when $\beta \to \infty$
and the network is an RGG. Our simulation shows this bound to be
$\mu_{{*}}^{(\beta\to \infty)} \approx 4.5$, in agreement with the simulation
result 
reported in \cite{Dall2002}. For $\beta=3$, the critical 
mean degree for percolation appears to be $\mu_{{*}}^{(\beta=3)} \approx3.1$
from the crossing point of the curves corresponding to different network sizes
in \fig{INn3n3_5n4n4_5n5mu0D2b3sOUTc}{}. 
\fig{INMsn4mu2mu3mu4mu5D2sOUTc}{} shows the size of the largest component
as function of $\beta$ for fixed values of $\mu$.  
Consistent with \fig{INn4mu0D2b0b1b3b10binfsOUTc}{}, we see that for $\mu 
< \mu_{{*}}^{(\beta\to \infty)} $ there is a maximum $\beta = \beta_{*}$ for the
existence of a giant component, 
while for $\mu> \mu_{{*}}^{(\beta\to \infty)}$ there is a giant component for
all values of $\beta$.

\begin{figure}[h!]
\centering
\includegraphics
[width=.6\textwidth]
{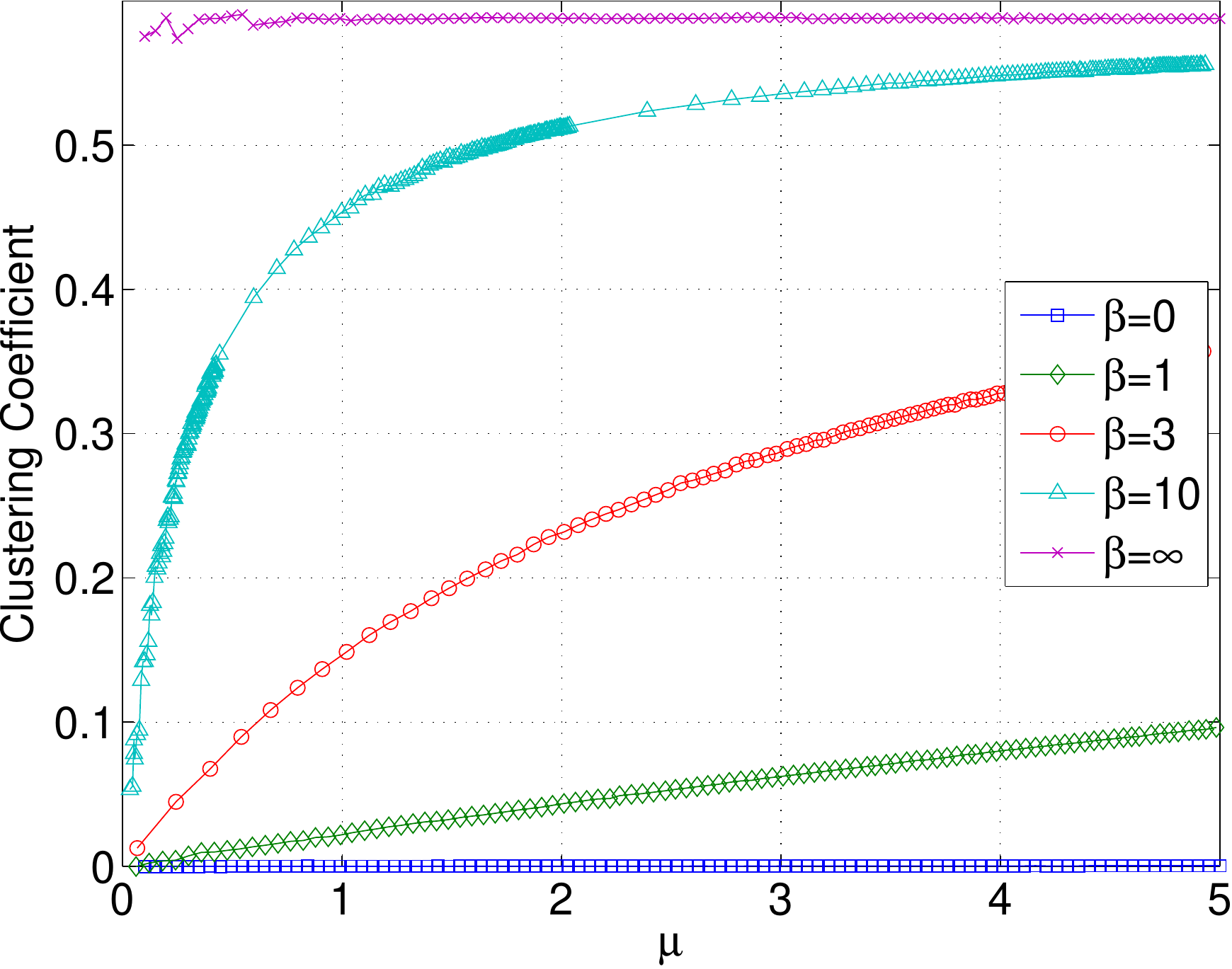}
 
\caption{Clustering coefficient as a function of $\mu$ for various values 
$\beta$. (The non-uniform distribution of data points in $\mu$ correspond to
uniform data points in $\kappa_{\beta
\mu}^{(D)}$.)}\label{INMsn4mu0D2b0b1b3b10binfsOUTC}
\end{figure}

For a given $\mu$, the clustering of vertices increases as $\beta$ increases,
achieving the maximum value as $\beta \to \infty$ (see \fig{INMsn4mu0D2b0b1b3b10binfsOUTC}{}). Substituting $D=2$ in \eqref{cow}, we find
that $C = 1 - 3\sqrt{3}/4\pi \approx 0.59$. 

\subsection{A model for a social network}
The unconstrained model can be used as a model of a social network where the
individuals have fixed opinions on $D$ number of issues. The parameter $\beta$
represents the tendency of individuals to befriend others of opinions similar 
to theirs, also known as homophily. The limitation in the number of active
social contacts an average person can maintain is represented by the fixed value
of the mean degree $\mu$. 
With the above interpretation of the parameters, the properties of the
unconstrained ESNM are compatible with those of real social networks. 
First, clustering, which is a central feature of any social network is easily
captured by the model (\fig{INMsn4mu0D2b0b1b3b10binfsOUTC}{}). 

Second, the absence of a giant component would imply a fragmented social network
(\fig{INn4mu0D2b0b1b3b10binfsOUTc}{}). So the critical mean degree
$\mu_*^{(\beta)}$ is the 
minimum number of friends that individuals need to make, so that a positive fraction of the social
network is connected. The stronger the preference of individuals to
connect to similar individuals (i.e., large $\beta$), the larger the number of 
friends they need to make (large $\mu$) to prevent disintegration. However, even
with a very high homophily, if the number of friends is at least
$\mu_{{*}}^{(\beta\to \infty)} \approx 4.5$, the social network is guaranteed to
be have a giant component.  

The probability of an edge from a vertex to its spatial neighbor decreases with the distance. Since the density of vertices is constant for $D=1$, but increasing with distance for $D \geq 2$, 
the corresponding edge length distributions have a maxima at zero and at a positive value, respectively.
For the social network, this implies that when there is
only a single issue on which opinions matter, 
the individuals mostly connect to others with opinions very close to theirs. On the other hand, when individuals choose their friends based on their opinions on multiple
issues, the highest concentration of friends is found at a positive opinion difference.

\comment{
For $D=1$, the distribution \eqref{crimea} of edge lengths is monotonically
decreasing. This means than when there is
only a single issue on which opinions matter, 
the individuals mostly connect to their closest spatial neighbors. However, for
$D \geq 2$, the distribution has a maxima (Fig. \ref{EdgeLengthDistribution}). 
So when individuals choose their friends based on their opinions on multiple
issues, most of the friends are located farther away.}

\section{The Connected ESNM}
\begin{figure}[h]
\centering
\subfigure[$\mu=2$]{\includegraphics
[width=.45\textwidth]
{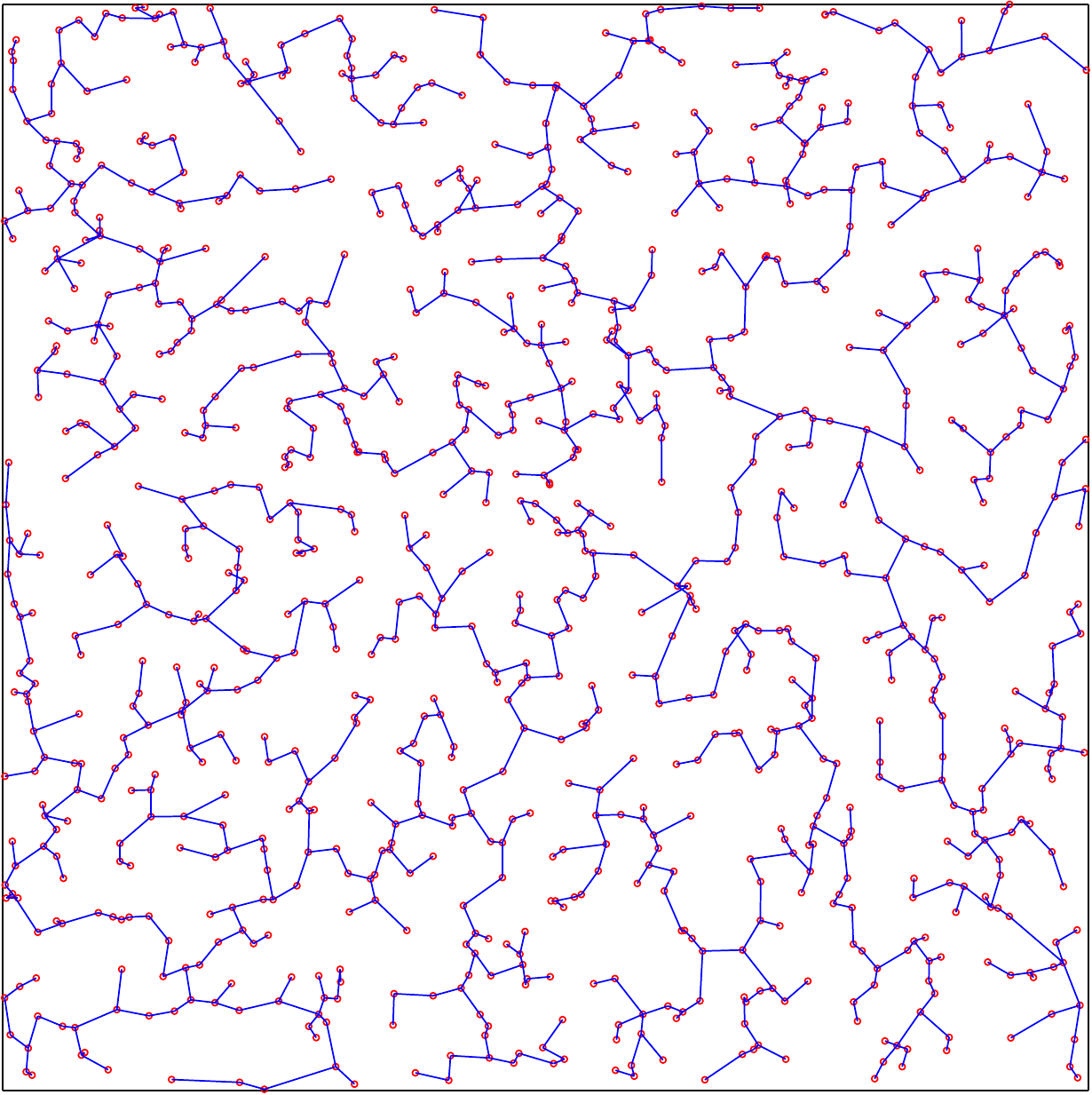}
}\,
\subfigure[$\mu=4$]{\includegraphics
[width=.45\textwidth]
{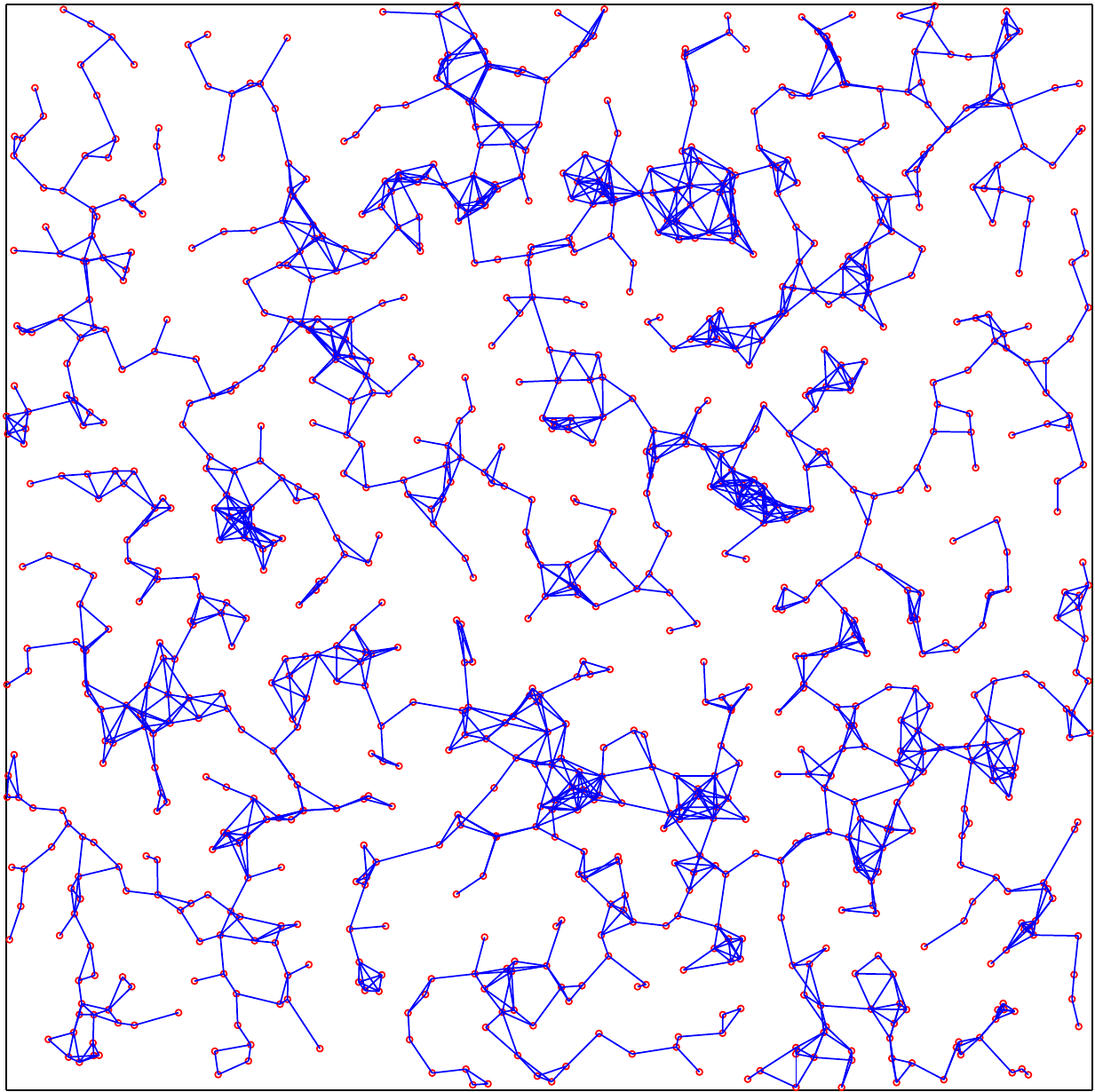} }
\caption{A realization of the almost optimized network ($\beta =10$)for two
different values of the mean degree. The red circles and blue lines correspond
to the vertices and edges respectively}
\label{INMon3mu2mu4D2b0k0sOUTpicture}
\end{figure}
We now consider our model with the constraint $\mathcal{T}$ that the network be
connected. Such a requirement is natural for many real world networks, for e.g.,
airline networks, road networks \cite{Barrat2005}. 
Although, we know the equilibrium distribution \eqref{eqdist} of the network, it
is difficult to proceed further analytically as we can no longer define an
equivalent percolation version of the model as we did for the unconstrained
model. 
The connectedness constraint makes the edges of the equilibrium network highly
correlated. Therefore we  study the connected ESNM purely by simulation. 
For simplicity, we will focus on two cases: when the parameter $\beta$ is zero
and when it takes a large value 10. 
We will refer to the $\beta=0$ equilibrium network as the \emph{Random Connected Network} or RCN($\mu$), 
and for reasons that will be elucidated in Section \ref{opti}, the 
$\beta=10$ network will be called the \emph{Almost Optimized Network} or AON($\mu$).

For the simulation we choose the dimension $D =2$. 
Since it is computationally costly to verify that the network remains connected, we will reduce our simulation size for the connected network to $n=10^3$
The initial network is formed by randomly ordering the vertices and adding $n-1$
edges to form a chain. 
The remaining $n \mu/2 - (n-1)$ edges are randomly chosen from the remaining
vertex pairs. 
We say that equilibrium has been reached when the mean edge length changes by
less that 0.5\% across time points separated by a large number (1000 times the
number of edges) of network update attempts. \fig{INMon3mu2mu4D2b0k0sOUTpicture}{} shows the network for two values of the mean
degree.

\subsection{Mean edge length, degree distribution, and clustering}
\begin{figure}[h]
\centering
\includegraphics
[width=.5\textwidth]
{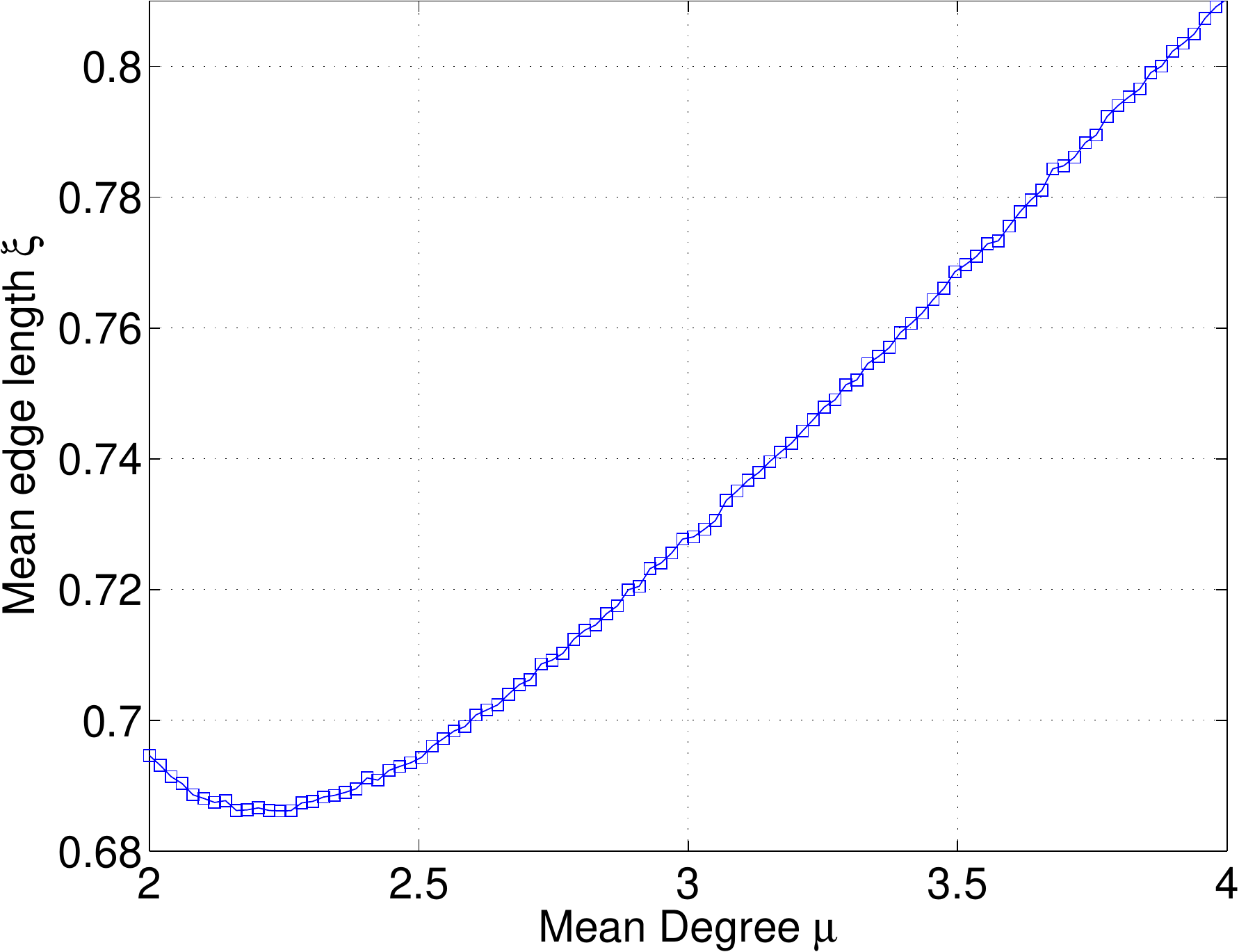}  
\caption{Mean edge length $\xi$ of the AON.\label{INMon3D2b10k0sOUTu}}
\end{figure}
As in the unconstrained case, the mean edge length blows up in the RCN
 for $n \to \infty$. The mean edge length as a function of
$\mu$ for the AON is shown in \fig{INMon3D2b10k0sOUTu}{}. 
It is interesting to note that that $\xi(\mu)$ is not 
monotone, but achieves a minima around $\mu=2.2$. 
As $\mu$ increases from 2 until about 2.5, the increase in the total length of
the network seems to be overcompensated by the increased flexibility in keeping
the network connected, resulting in short edge lengths.

\begin{figure}[h]
\centering
\subfigure[RCN]{\includegraphics
[width=.45\textwidth]
{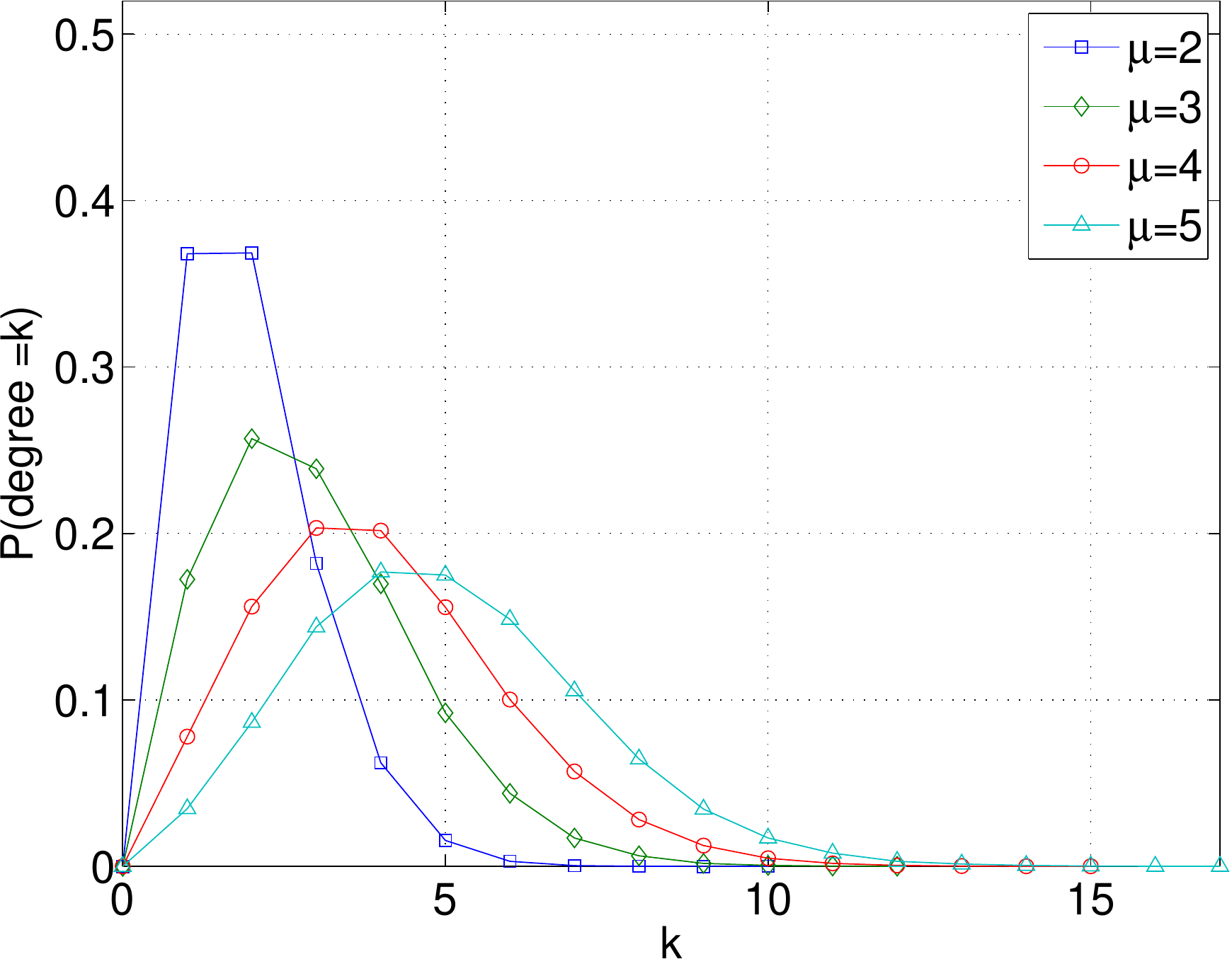} }\,
\subfigure[AON]{\includegraphics
[width=.45\textwidth]
{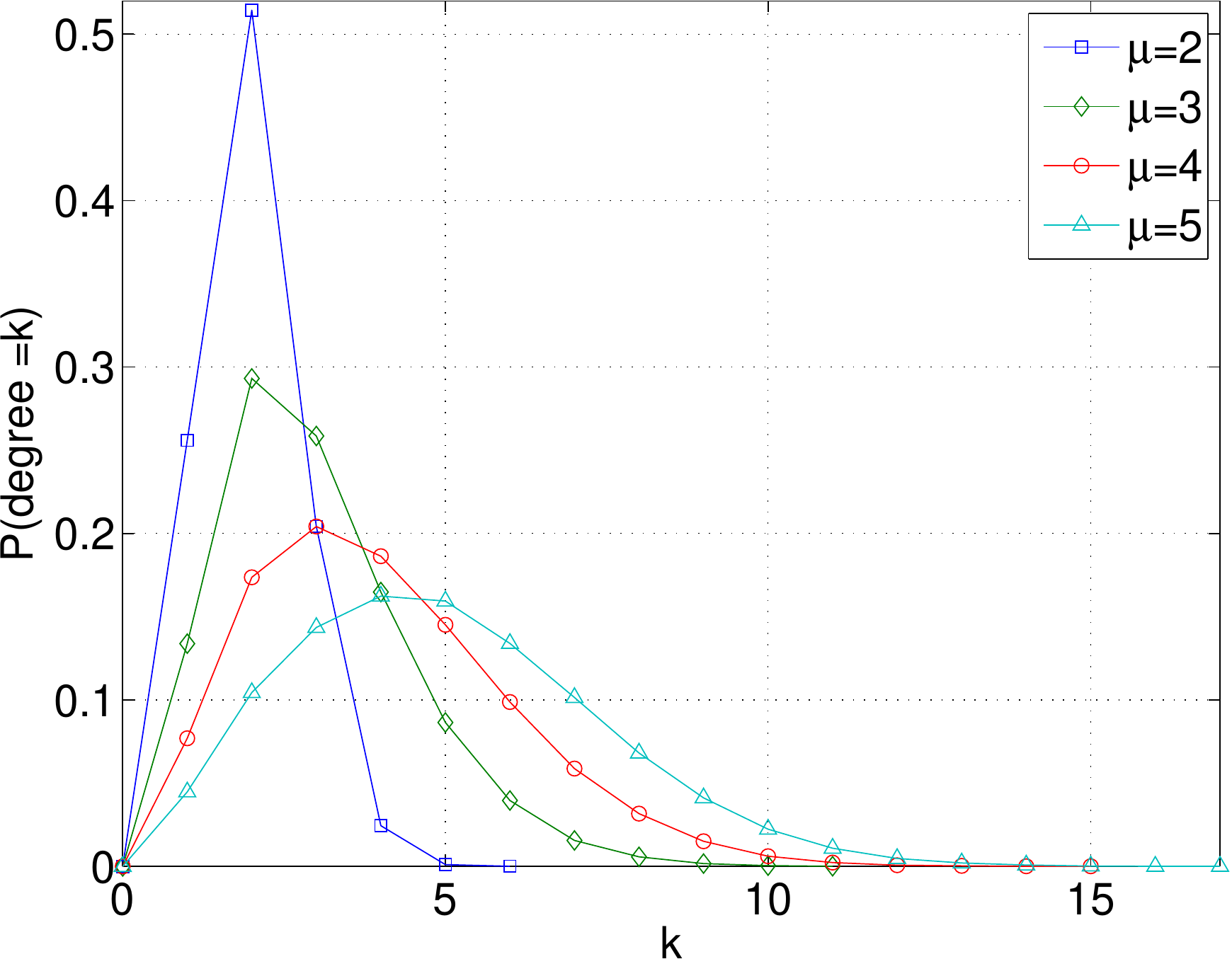} }
\caption{Degree distribution of the random connected and optimized networks.}
\label{INMon3mu2mu3mu4mu5D2b0b10k0sOUTDegreeDist}
\end{figure}
In contrast to the unconstrained network, the degree distribution of the
connected network does not  appear to be Poisson for any value of $\beta$, as
seen in \fig{INMon3mu2mu3mu4mu5D2b0b10k0sOUTDegreeDist}{}. 
The distribution however is still peaked around the mean with a thin tail. For
$\mu=2$, the AON has a markedly higher peak at $2$ 
than the RCN\comment{, indicating the presence of long chains}. 

\begin{figure}[h]
\centering
\includegraphics
[width=.5\textwidth]
{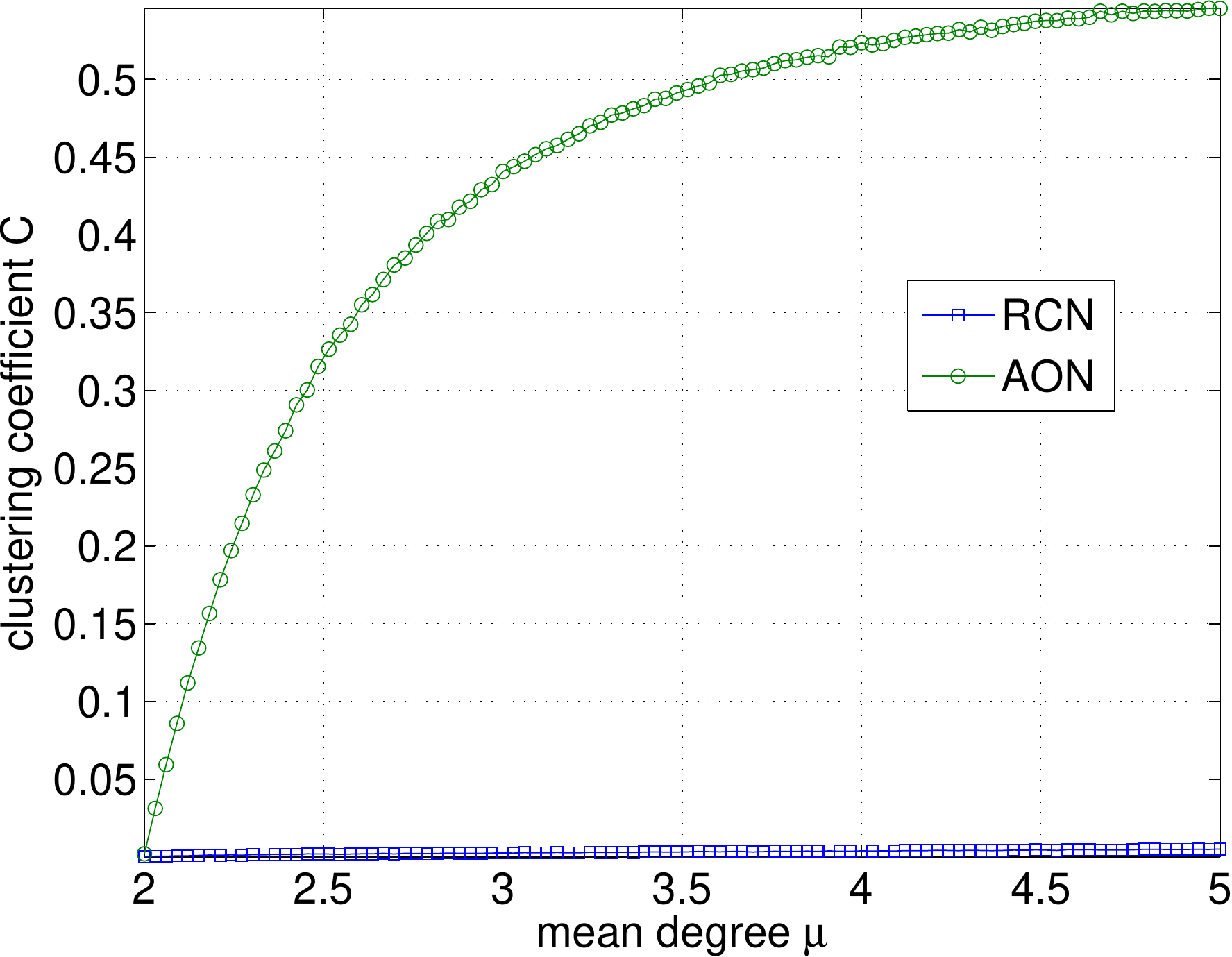}  
\caption{Clustering coefficient as a function of the mean degree
$\mu$.}\label{INMon3D2b0b10k0sOUTC}
\end{figure}
For the RCN, since the spatial locations are unimportant, it is natural that the
clustering coefficient vanishes (as $n \to \infty$). 
The AON, on the contrary, has high clustering as shown in \fig{INMon3D2b0b10k0sOUTC}{}.

\subsection{The $\beta \to \infty$ model as an optimization process}
\label{opti}
In the $\beta \to \infty$ limit, the connected ESNM may be viewed as a
stochastic algorithm (although not a very efficient one) to solve the following
optimization problem:
Given a collection of $n$ points uniformly distributed in $\mathcal{V}_{nD}$ and
$\mu \geq 2 - 2/n$ (i.e., the number of edges is at least the minimum $n-1$
needed to connect $n$ vertices), 
find the \emph{connected} spatial network $G_*$ with mean degree $\mu$ that has
the lowest total length, i.e., find
\begin{align}
 G_*(\mu) =  \underset{G \in \mathcal{G}_c(V_{nD},\mu)}{\arg\max} H(G)\,,
\end{align} 
where $\mathcal{G}_c(V_{nD},\mu)$ is the set of connected networks with vertex
set $V_{nD}$ and $n \mu/2$ edges

Now, consider the general problem of finding an ``efficient'' network over a
given collection of points. Obviously, the application one is interested in
determines the optimization metric \cite{Thai2012}. 
A simple and very popular optimal network is the minimum
spanning tree abbreviated as MST (see \cite{Graham1985} for a history and
\cite{Prim1957} for a classic algorithm). 
Here the quantity that is minimized is the total length, or equivalently, the
``wiring cost'' of the network.
\comment{For the same optimization problem, if one is allowed to introduce
additional points to the existing collection, then the wiring cost could be
further reduced resulting in a Steiner tree \cite{Gilbert1968}. }
$G_*(\mu)$ is very similar to the MST with the notable exception that it is not
a tree for $\mu\geq 2$. Indeed, $G_*(\mu = 2 - 2/n)$ is the MST.

However, one could potentially be concerned about other aspects of the network
in addition to its wiring cost, and a tree may no longer be a good option. 
For example, Aldous \cite{Aldous2008b, Aldous2010} sought networks which in
addition to minimizing the wiring cost 
also has short routes, i.e., the route distance between any pair of vertices is
close to their spatial distance. 
He quantified this property by defining the \emph{route factor} $R(x,y)$ between
two vertices $x$ and $y$ as 
\begin{align}
 R(x,y) = \frac{r(x,y)}{|x-y|} -1\,,
\end{align}
where $r(x,y)$ is the \emph{route distance} or the length of the shortest route between $x$ and $y$.
The route factor defined for a single vertex pair can then be averaged over all
vertex pairs to arrive at 
a useful statistic for the network -- the \emph{mean route factor} $R$.
\comment{
Gastner and Newman defined edge weights to account for the route distance
\cite{Gastner2006} and hop distance \cite{Gastner2006a} between vertices. }
Gastner and Newman \cite{Gastner2006} studied a growth model for spatial networks, where given a $V_{n2}$ (i.e., vertices distributed uniformly in a square) 
with a designated ``root'' vertex, a connected cluster is grown by sequentially adding edges to vertices outside the cluster; 
the edges are chosen according to a greedy optimization criterion that minimizes a linear combination 
of the new edge length, and the route factor between the new and the root vertices.

One may also want the network to be robust to random failures of its edges. 
\comment{or vertices \cite{Paul2004}.
If one is only concerned about edge failures, then} 
One way to test this kind of robustness of a connected network is by randomly
removing edges \cite{He2009} and 
noting the size of the largest component of the resulting network. Specifically,
for a connected network of mean degree $\mu$ we look at the fraction
$\rho_\mu(\mu')$ 
of vertices in the largest component when the edge removal leads to a network of
mean degree $\mu'$. 
A robust network should retain a large fraction of its vertices in its largest
component when $\mu'$ decreases from $\mu$; in other words 
the function $\rho_\mu(\mu')$ should be concave downwards for a sizeable region
near $\mu'=\mu$. 
We thus quantify the robustness of the network by the inflection point
$\tilde{\mu}(\mu)$ of the $\rho_\mu(\cdot)$ curve. Note that lower $\tilde{\mu}(\mu)$ means more robust. 
The $\rho_\mu(\cdot)$ curve may always be convex indicating the lack of
robustness of the network;  
therefore, for a collection of networks parametrized by their mean degrees
$\mu$, we define the critical mean degree $\mu_*$ for robustness as the smallest
$\mu$ for which there exists an inflection point.

Thus, similar to \cite{Tero2010}, we characterize the efficiency of a given
network of mean degree $\mu$, by three statistics: 
the scaled total edge length $\chi = H(G)/n^{1+1/D}$, the route factor
$R$, and $\tilde{\mu}$. The smallness of all these network statistics is
desirable for an efficient network. 
How does $G_*(\mu)$ fare in these measures of efficiency? 
In order to get an approximation to $G_*(\mu)$, we perform simulations using
$\beta=10$ and term the equilibrium network as the \emph{almost optimized
network} or AON($\mu$). 
Since the equilibrium of the $\beta=0$ connected model is uniformly drawn from
the set $\mathcal{G}_c(V_{nD},\mu)$ without regard for the edge length, it can
be viewed as a null model for comparison with the AON, and we will 
refer to it as the Random Connected Network or RCN($\mu$). 

\begin{figure}[h]
\centering\comment{
\subfigure[]{\includegraphics
[width=.47\textwidth]
{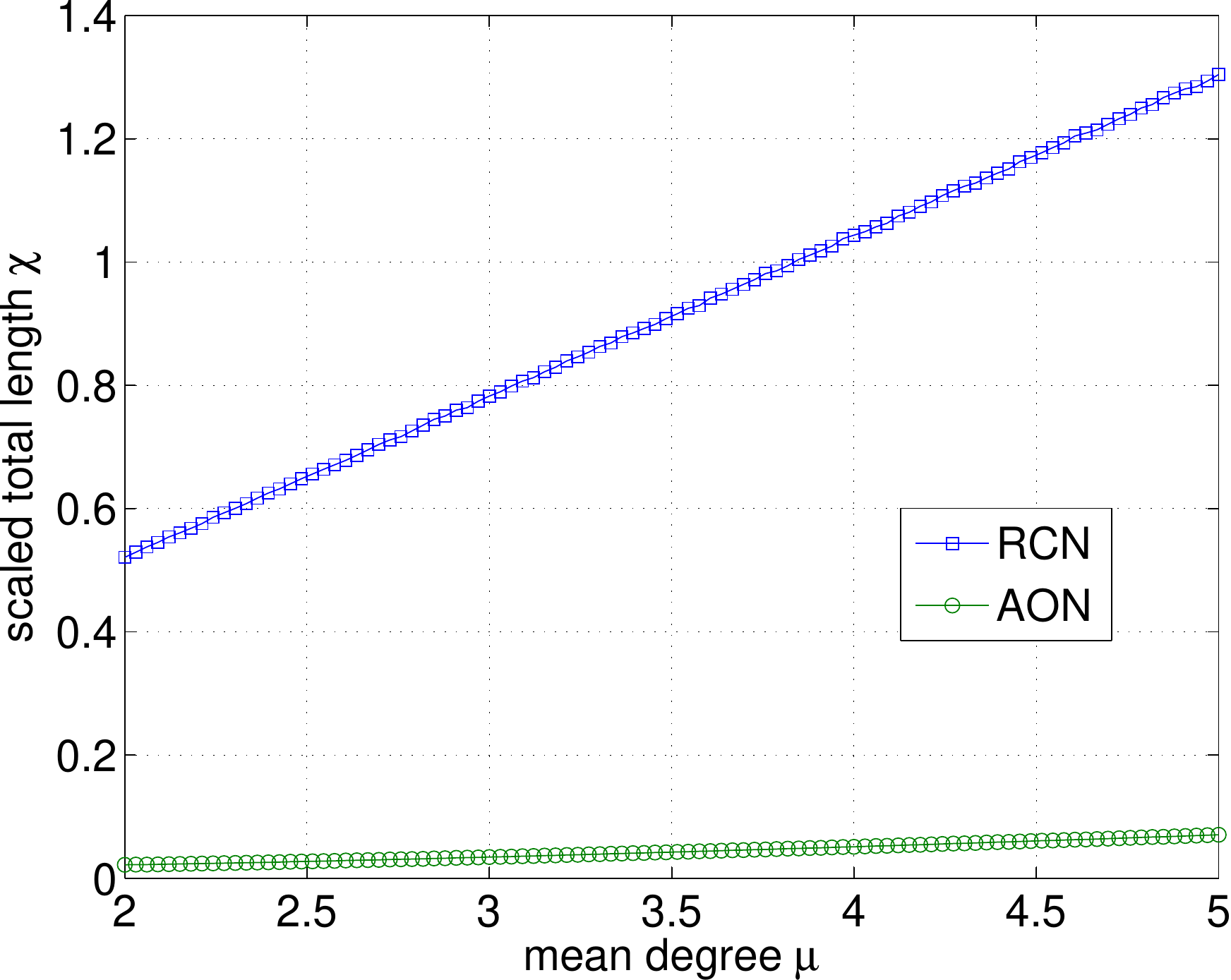} \label{INMon3D2b0b10k0sOUTchimu}}\,
\subfigure[]{\includegraphics
[width=.47\textwidth]
{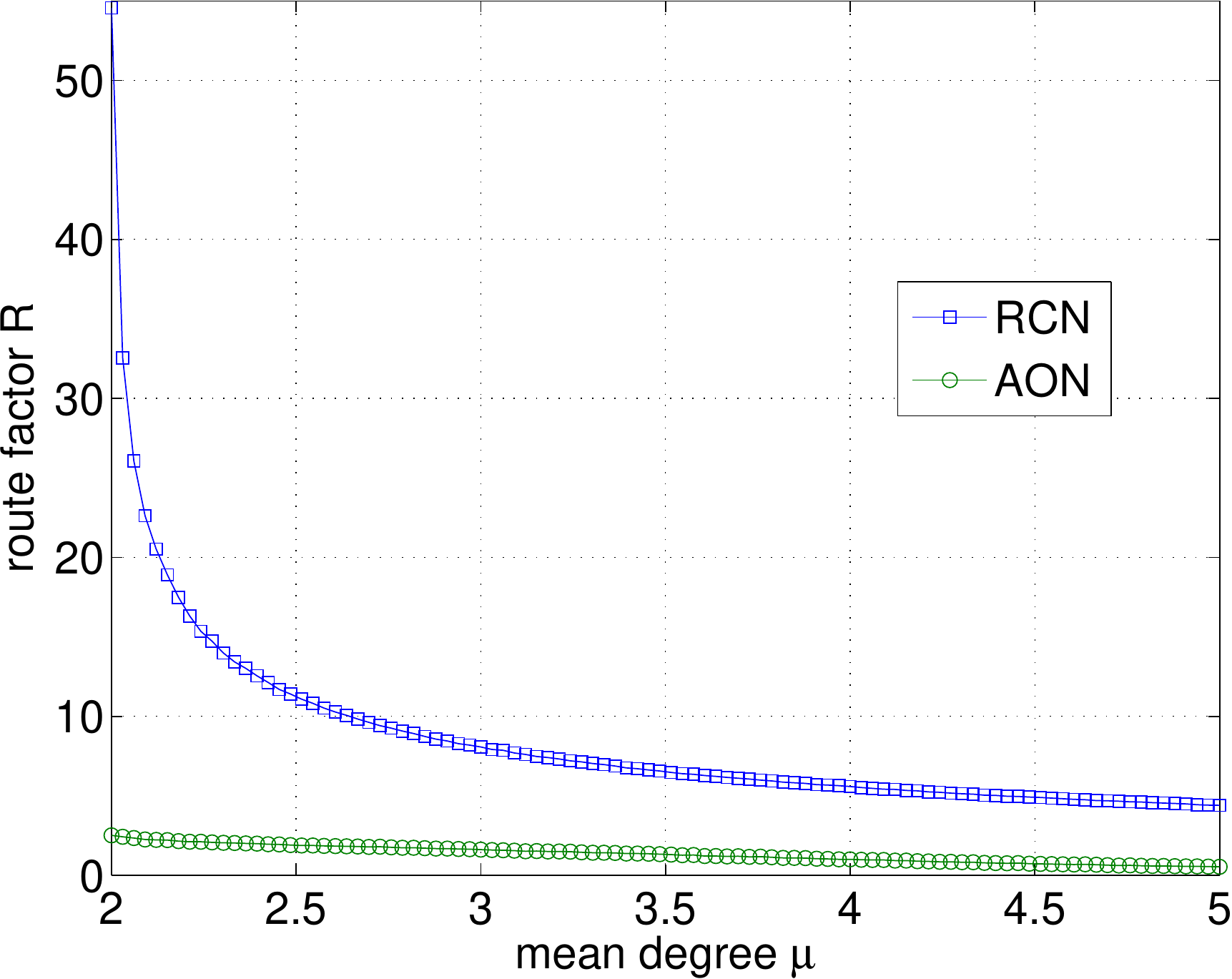} \label{INMon3D2b0b10k0sOUTR}}\\}
\subfigure[RCN]{\includegraphics
[width=.47\textwidth]
{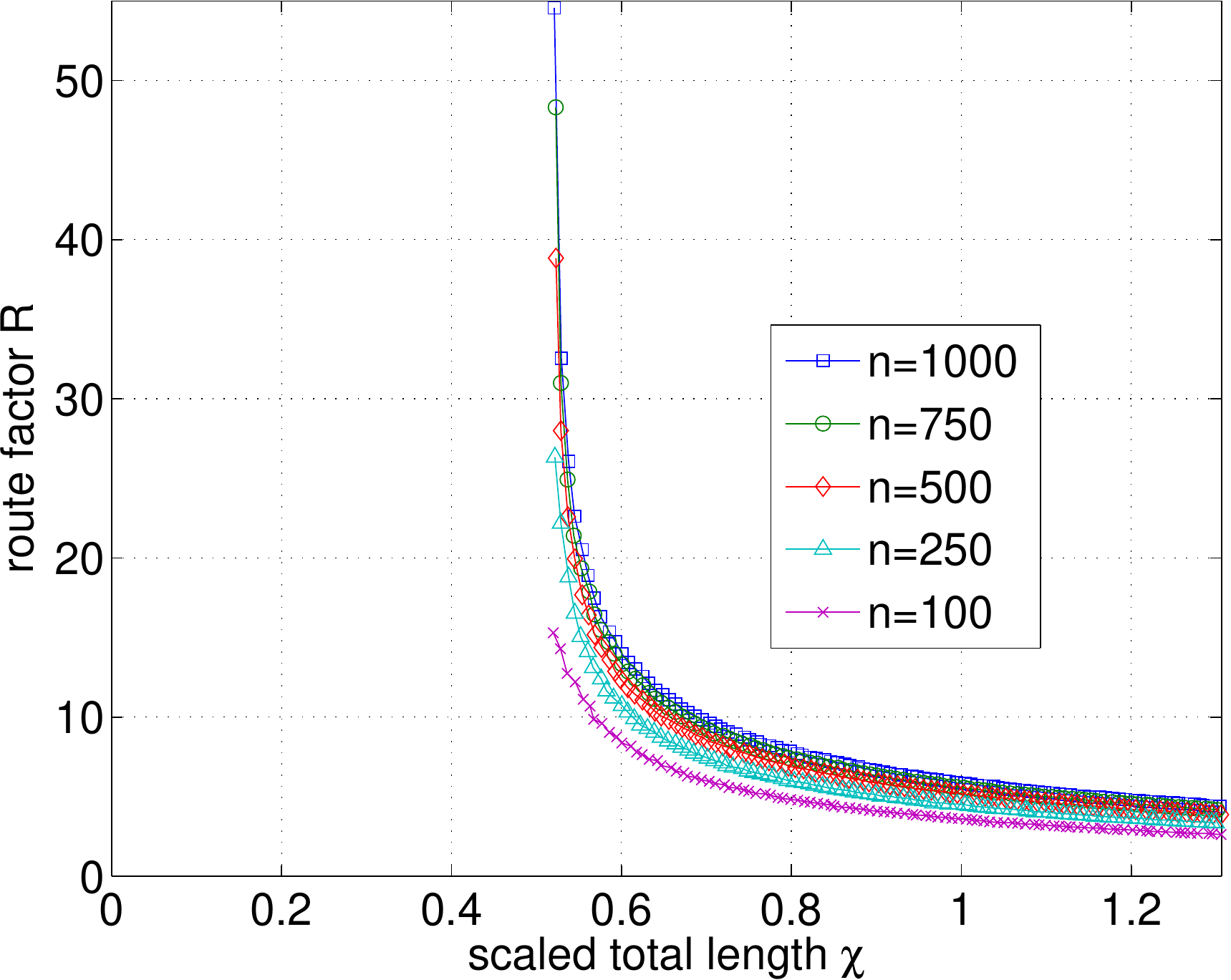}
\label{INMon3n2_8n2_7n2_4n2D2b0k0sOUTchiR}}\,
\subfigure[AON]{\includegraphics
[width=.47\textwidth]
{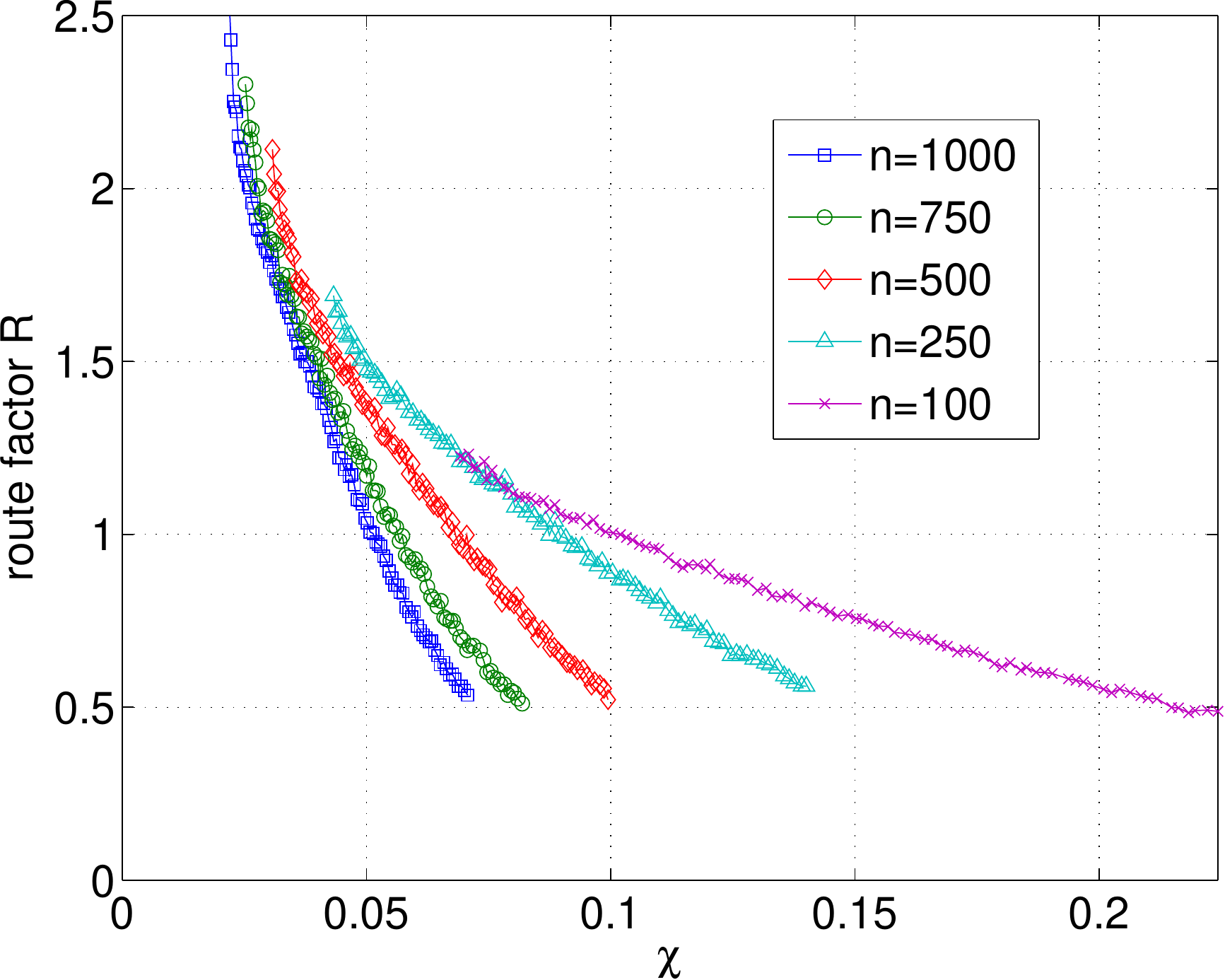}
\label{INMon3n2_8n2_7n2_4n2D2blargek0sOUTchiR_special}}
\caption{The route factor $R$ versus the scaled total length $\chi$ for the RCN and the AON.}
\end{figure}
\comment{Although intuitive, it is not completely obvious that the route factor, and the scaled total length will monotonically 
decrease and increase, respectively with the mean degree, as seen in Fig.~\ref{INMon3D2b0b10k0sOUTchimu} and \ref{INMon3D2b0b10k0sOUTR}.}
The two opposing statistics --  route factor and scaled total length, are plotted against each other in Fig.~\ref{INMon3n2_8n2_7n2_4n2D2b0k0sOUTchiR} and \ref{INMon3n2_8n2_7n2_4n2D2blargek0sOUTchiR_special}  
 to get convex ``efficiency curves'', which show that
the AON is significantly 
more efficient than the RCN if we only take $\chi$ and $R$ into account. 
The finite size scaling shows that the route factor diverges at $\chi \approx 0.52$ for the RCN corresponding to the random connected tree, 
and at $\chi \approx 0.02$ for the AON corresponding to the MST.

\begin{figure}[h]
\centering
\subfigure[]{\includegraphics
[width=.45\textwidth]
{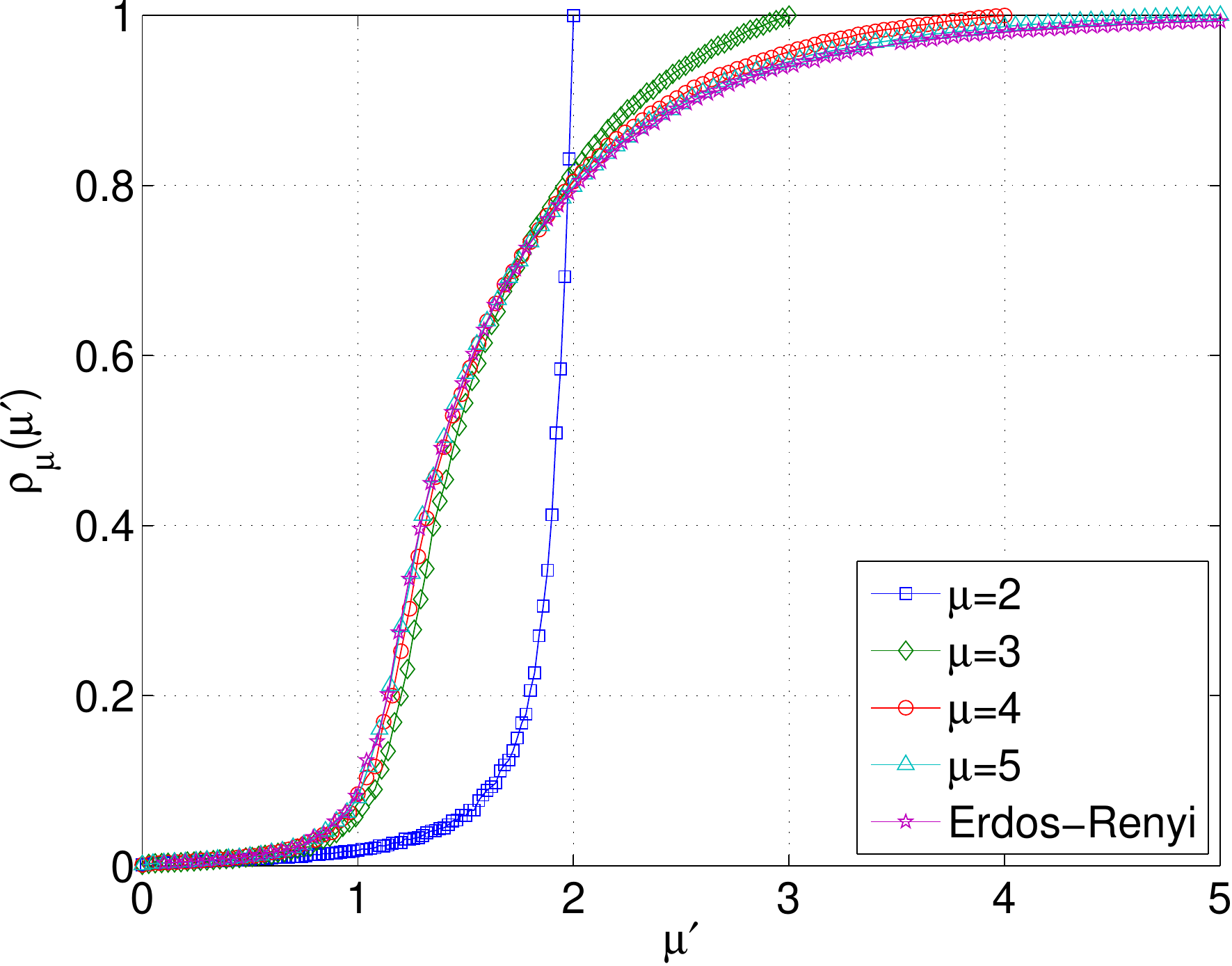}\label{INMon3mu2mu3mu4mu5D2b0k0sOUTRobustness} }\,
\subfigure[]{\includegraphics
[width=.455\textwidth]
{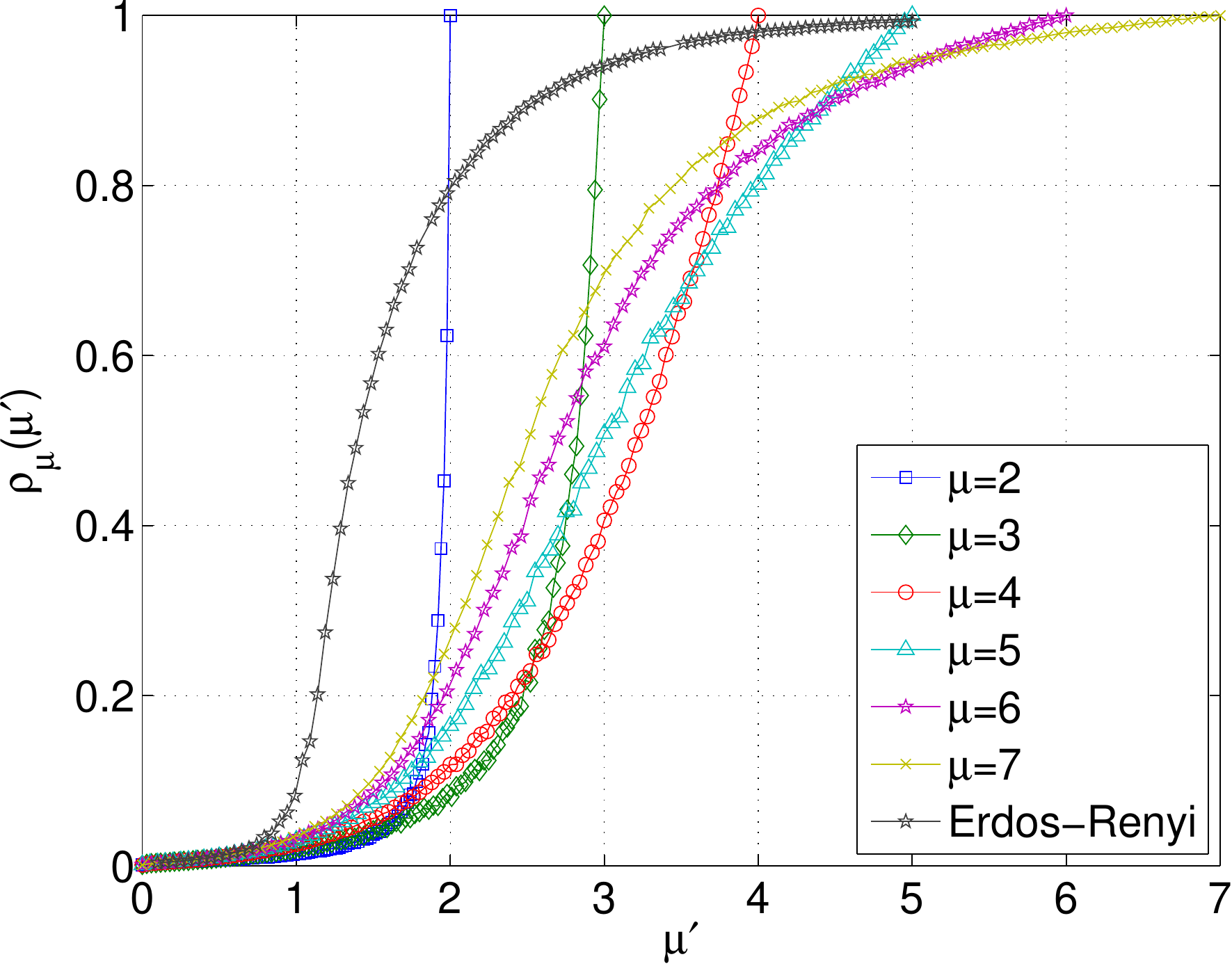} \label{INMon3mu2mu3mu4mu5mu6D2b10k0sOUTRobustness}}
\caption{Robustness of the (a) RCN and (b) AON.}
\label{INMon3mu2mu3mu4mu5D2b0b10k0sOUTRobustness}
\end{figure}

However, in terms of robustness to random edge failures, the RCN with its
abundance of long range 
connections performs better as shown in \fig{INMon3mu2mu3mu4mu5D2b0b10k0sOUTRobustness}{}. 
For comparison, we also include  the fraction of vertices in the largest
component of an Erd\H{o}s-R\'{e}nyi
random graph (note than an ER graph with edges removed at random is just another
ER graph with a lower mean degree). 
While it is difficult to precisely locate the inflection point
$\tilde{\mu}(\mu)$ in these curves, it is easy to see that 
it decreases with increase in the mean degree $\mu$, i.e., as one would expect,
more edges make the network more robust. 
In \fig{INMon3mu2mu3mu4mu5D2b0k0sOUTRobustness}{}, 
the $\mu =3,4,5$ curves are almost indistinguishable from the ER curve and show
the percolation 
transition close to $\mu'=1$, indicating that they are very similar to
Erd\H{o}s-R\'{e}nyi networks. However, the behavior of the Almost Optimized
Networks as seen in 
\fig{INMon3mu2mu3mu4mu5mu6D2b10k0sOUTRobustness}{} is quite different.
It can be inferred from \fig{INMon3mu2mu3mu4mu5D2b0b10k0sOUTRobustness}{}
that the critical mean density $\mu_*=2$ for the RCN, and $ 4<\mu_*<5$ for the
AON. 

\begin{figure}[h]
\centering
\includegraphics
[width=.5\textwidth]
{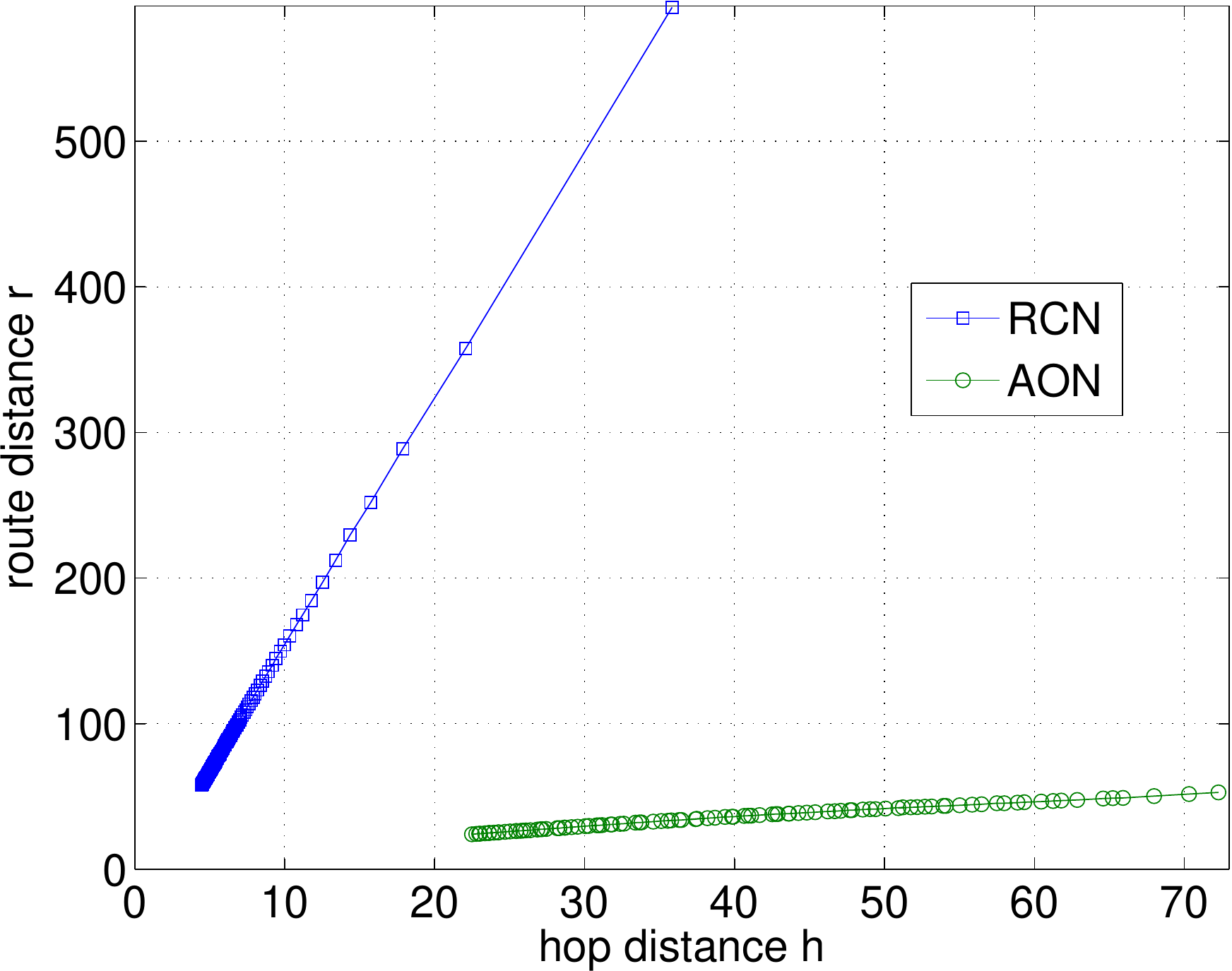}  
\caption{Mean route distance $r$ versus mean hop distance $h$ for the RCN and AON.}\label{INMon3D2b0b10k0sOUThr}
\comment{
\centering
\subfigure[\comment{Hop distance $h$}]{\includegraphics
[width=.45\textwidth]
{INMon3D2b0b10k0sOUTh.pdf}
\label{INMon3D2b0b10k0sOUTh}}\,
\subfigure[\comment{Route distance $r$}]{\includegraphics
[width=.45\textwidth]
{INMon3D2b0b10k0sOUTr.pdf}
\label{INMon3D2b0b10k0sOUTr}}
\caption{Hop distance $h$ and route distance $r$ as a function of the mean
degree $\mu$ for the RCN and AON.}}
\end{figure}
The mean route distance $r$ is lower for the
AON than for the RCN; however, in achieving a lower $r$, the AON gets a higher mean hop distance $h$
(\fig{INMon3D2b0b10k0sOUThr}{}). 

\subsection{Testing the model on real data}
In this section, we apply the connected ESN model on two sets of data, to gain
some insight into the applicability of the model. 

\begin{figure}[h]
\centering
\subfigure[]{\includegraphics
[width=.47\textwidth]
{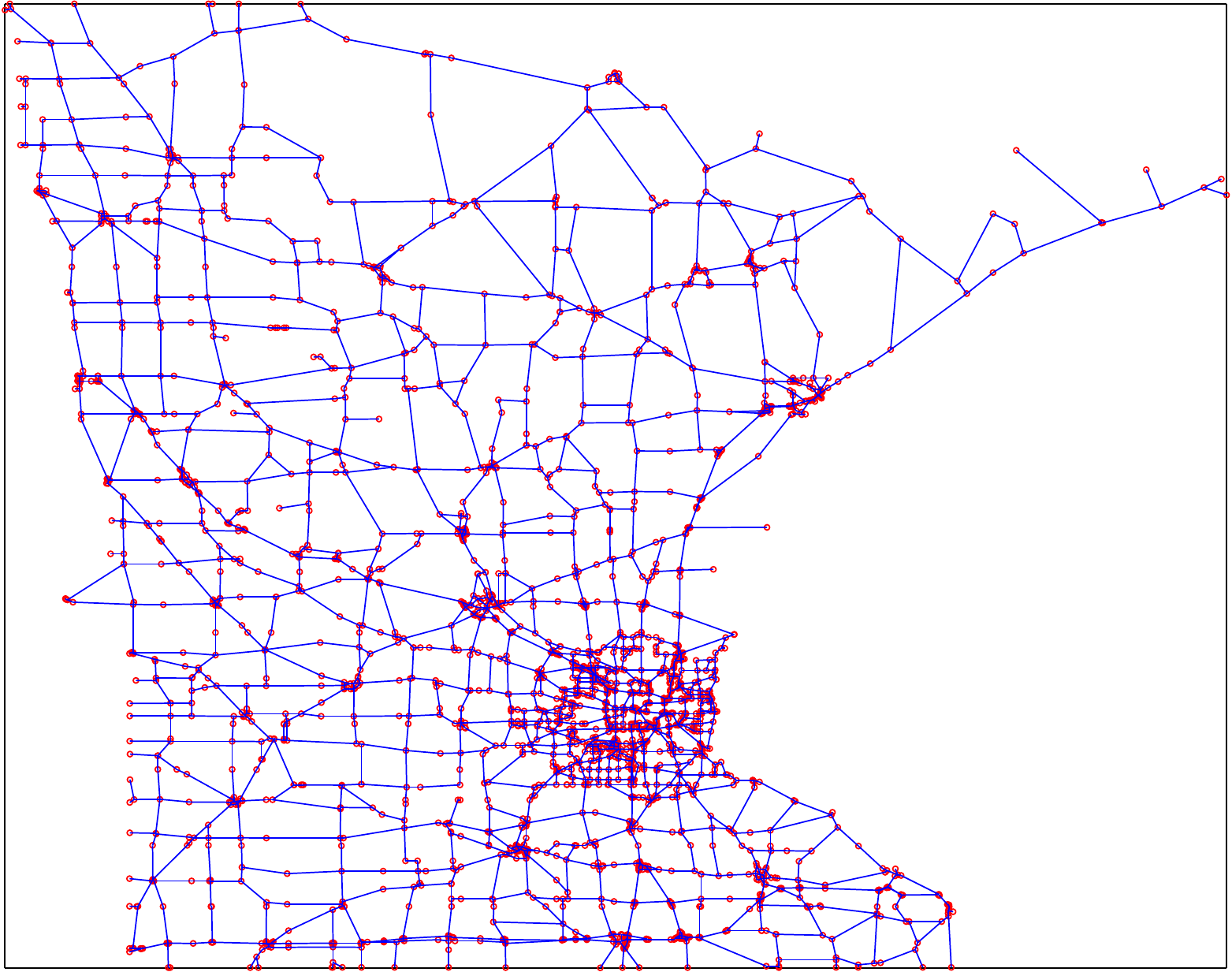}
\label{datapic}}\,
\subfigure[]{\includegraphics
[width=.47\textwidth]
{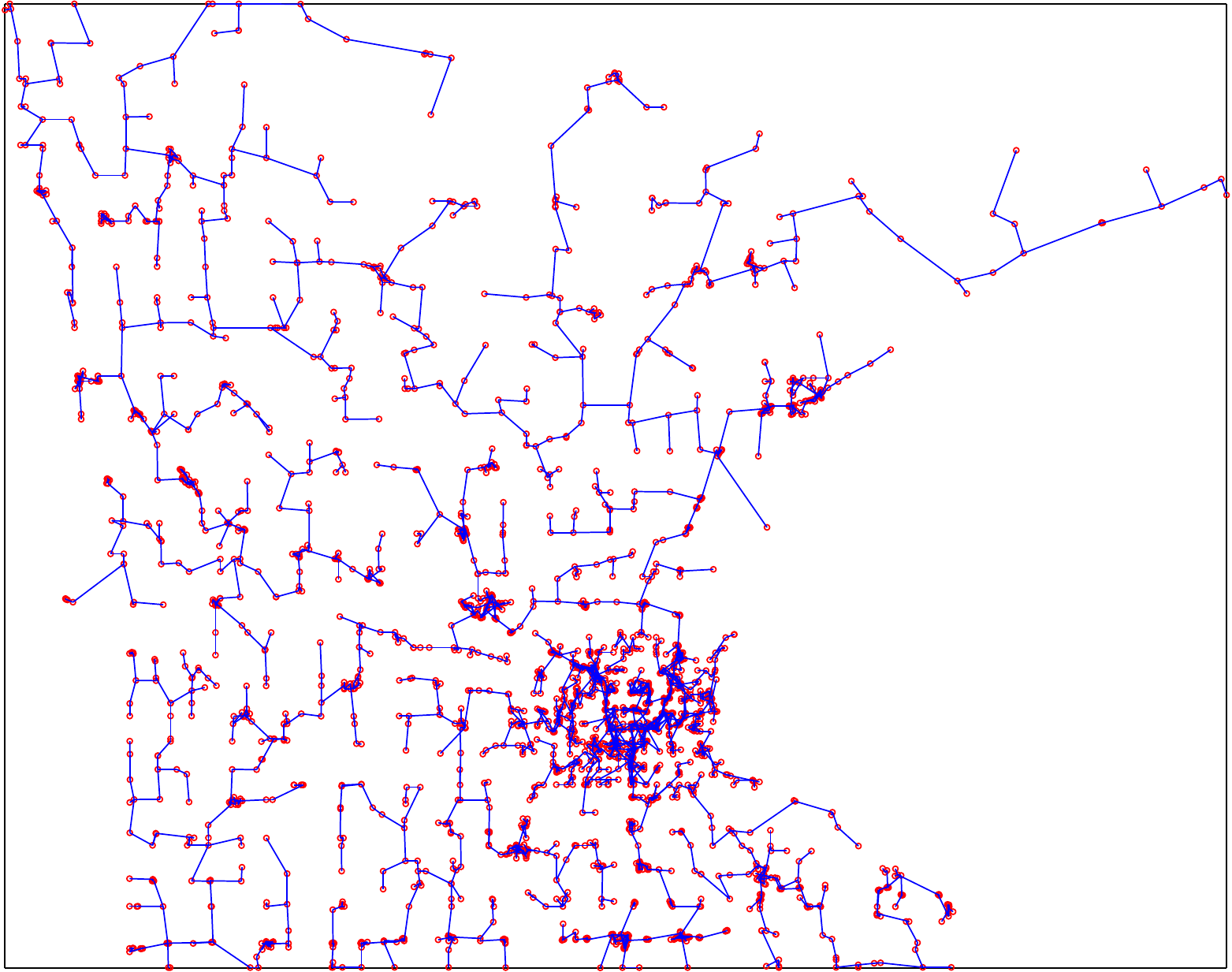} \label{simulpic}}
\caption{Pictures generated using the Minnesota road network data -- (a) the
actual network, and (b) the simulated network with $\beta=50$. \label{roadpics}}
\end{figure}

\begin{table}[h]
\centering
\begin{tabular}{|r|c|c|c|}
\hline
 & Data &  Simulation\\\hline
mean edge length $\xi$&  0.0683 &  0.0517  \\
clustering $C$ & 0.0280 &  0.114\\
hop distance $h $ &  80.0&   76.8 \\
route distance $r$ & 6.10&  5.76	\\
route factor $R$ & 1.79 &  1.64\\
\hline
\end{tabular}
\caption{Comparison of various statistics of the actual and
simulated networks. \label{roadstats}}
\end{table}

\begin{figure}[h]
\centering
\subfigure[]{\includegraphics
[width=.47\textwidth]
{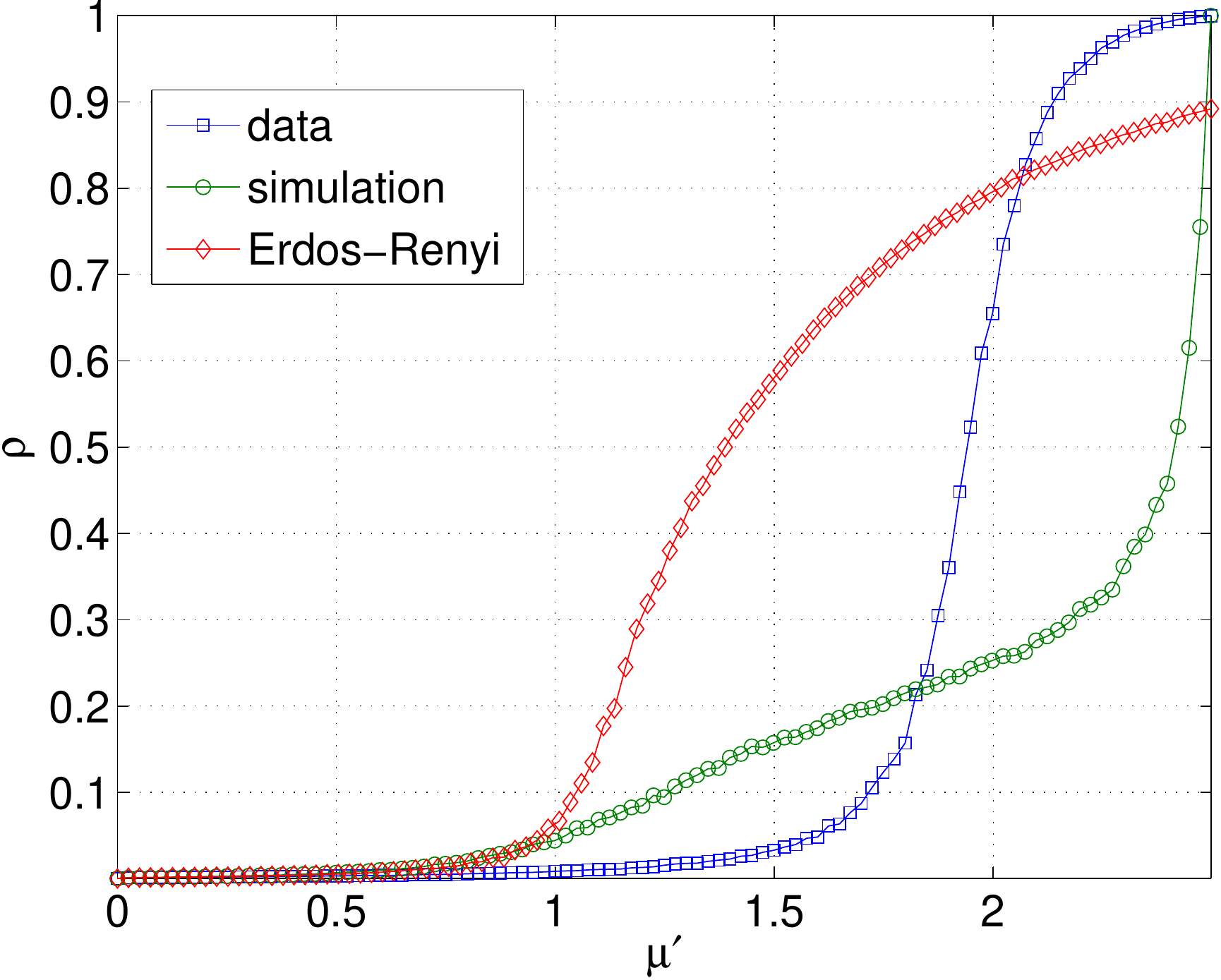} \label{INMinnesotamu0mu2_49791b50OUTRobustness}}
\subfigure[]{\includegraphics
[width=.47\textwidth]
{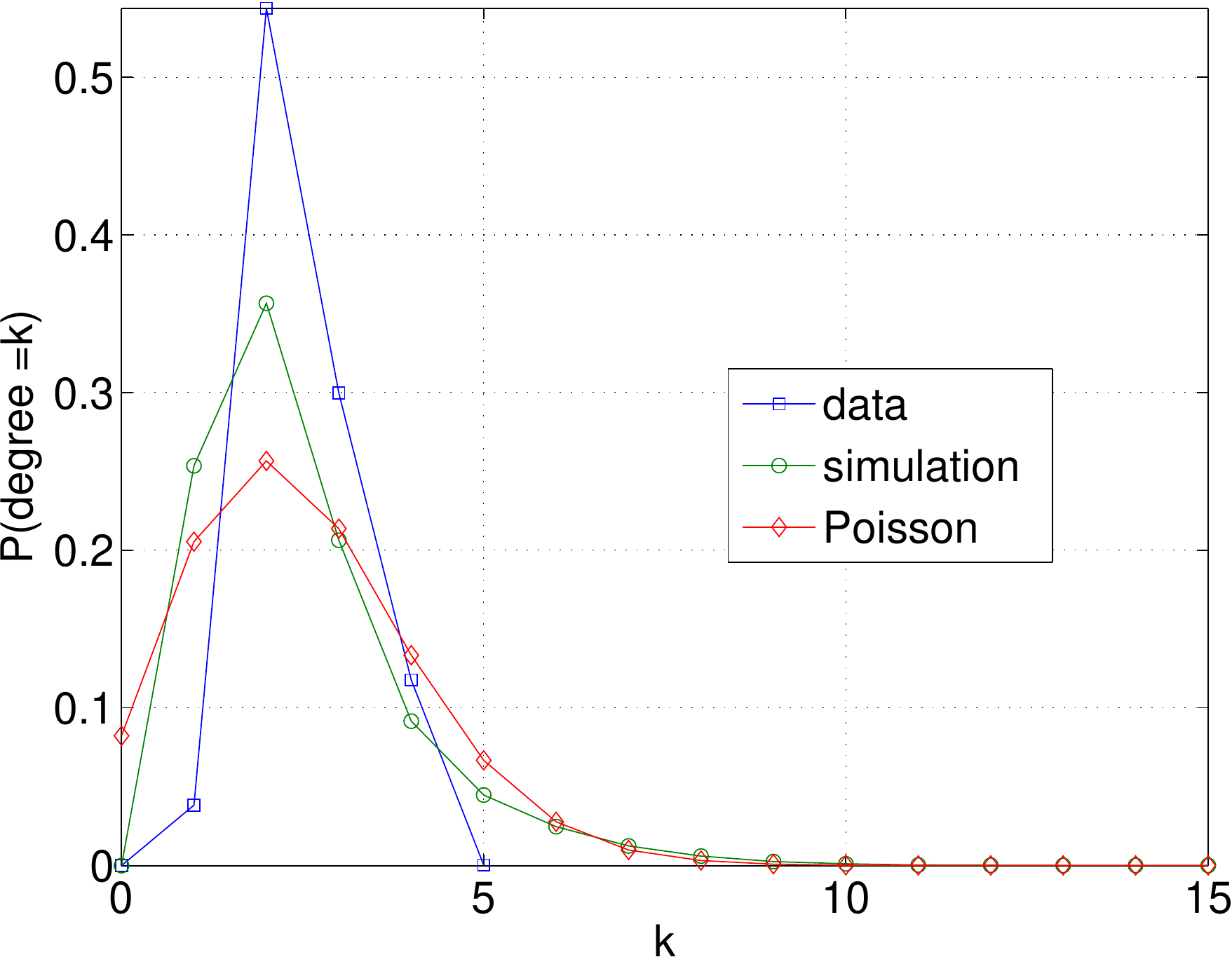} \label{INMinnesotamu0mu2_49791b50OUTDegreeDist}}\,
\caption{Comparison of (a) robustness and (b) degree distribution, of the
actual, simulated, and \ER networks.}
\end{figure}

Our first data set is about the network of roads in the US state of Minnesota,
obtained from \cite{minnesota} . 
There are $n=2635$ vertices in this network which correspond to the
intersections of the roads. 
The mean degree is $\mu \approx 2.5$.
\comment{$\mu = 2.49791$.} 
To obtain the simulated network, we run the connected ESNM on the vertex set of
the actual network, with the same mean degree and a large $\beta =50$.

In \fig{roadpics}{} we see that the actual and simulated networks look very
different. 
While the actual network has a grid like structure almost throughout, the
simulated network is tree like for the most part, expect for the small region 
(which corresponds to the capital city of Minneapolis) with a very 
high density of vertices. Table \ref{roadstats} compares the two network using
the statistics we used earlier, and we find that 
the simulated network performs better than the actual network on all of them. 
Specifically, the hop and route distances, and the route factor, which are all
measures of the ease of traversing the network, are marginally lower. 
Also for the simulated network, the construction cost of the roads measured by
the mean edge length is slightly lower, while the clustering 
coefficient is significantly higher\comment{ -- a desirable feature}. 

So does this mean that the simulated network is the more ``efficient'' and
``better'' network?
It does not seem likely that people 
living along the border with Canada would agree. 
In the ``optimized'' network they often have to go large distances to get to a nearby town, 
and even if they are driving to Minneapolis, they have a much longer route.
\comment{
Note that our algorithm look at distances between randomly chosen vertices, 
which is not weighted by the population density.   

First, notice that in the simulated network, the 
odds are significantly stacked against people living outside Minneapolis; for
instance, the journey from the north west corner to the capital city is a very
length one. 
In other words, the network design pretty much ignores everything outside the
capital region.} Second, how robust are the two networks to edge failures?
\fig{INMinnesotamu0mu2_49791b50OUTRobustness}{} shows that the 
simulated network is extremely fragile compared to the actual network; a loss of
less than 8\% of the edges is enough to bring the size of the largest component
down to a third of the network size. 
The actual network, on the contrary, is robust (by our earlier definition) with
a $\tilde{\mu} \approx 1.9$. \fig{INMinnesotamu0mu2_49791b50OUTDegreeDist}{} shows 
that in the actual network, a large fraction 
of the intersections are created by two roads, and no intersection is made of
more than four roads -- both of which are unsurprising. 
The simulated graph, while having a peak at 2, has unrealistic 9-road
intersections in the capital region.

The undesirable topology of the simulated network can be attributed mainly to the highly non-uniform
distribution of vertices in the graph (recall that our model assumes a uniform
distribution of the vertices). 
Our connected ESNM allocates a disproportionate amount of edges to regions  of
high vertex density. One may also argue about the quality of the statistics we
used; 
specifically, in practical applications, extreme values of the hop and route
distances and route factor are perhaps more relevant than their averages. 
Nevertheless, our robustness measure seems to be a reliable statistic for most
cases.

\begin{figure}[h]
\centering
\subfigure[$\mu=2$]{\includegraphics
[width=.47\textwidth]
{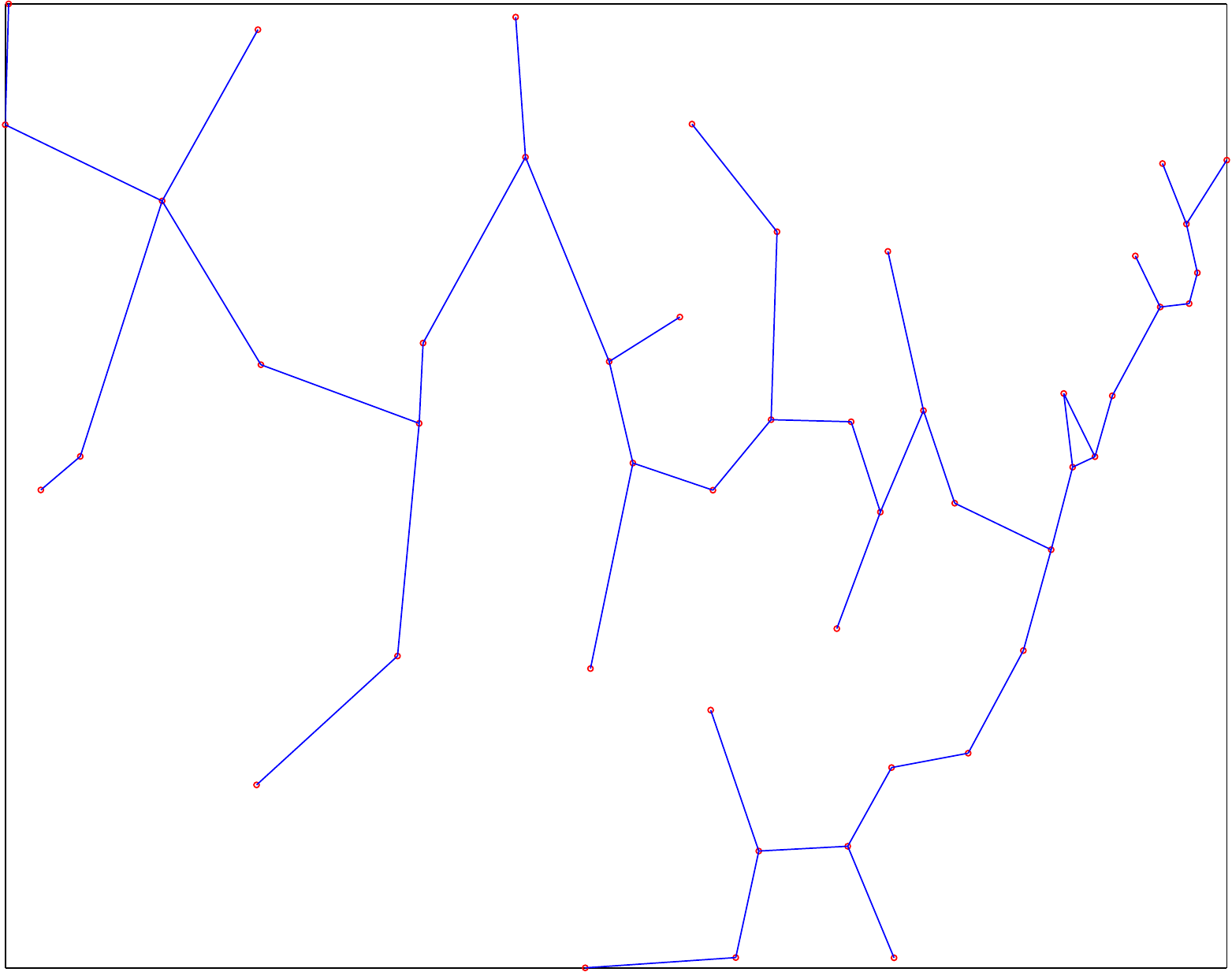}
\label{INUScapitalsxymu2b50OUTpicture}}\,
\subfigure[$\mu=3$]{\includegraphics
[width=.47\textwidth]
{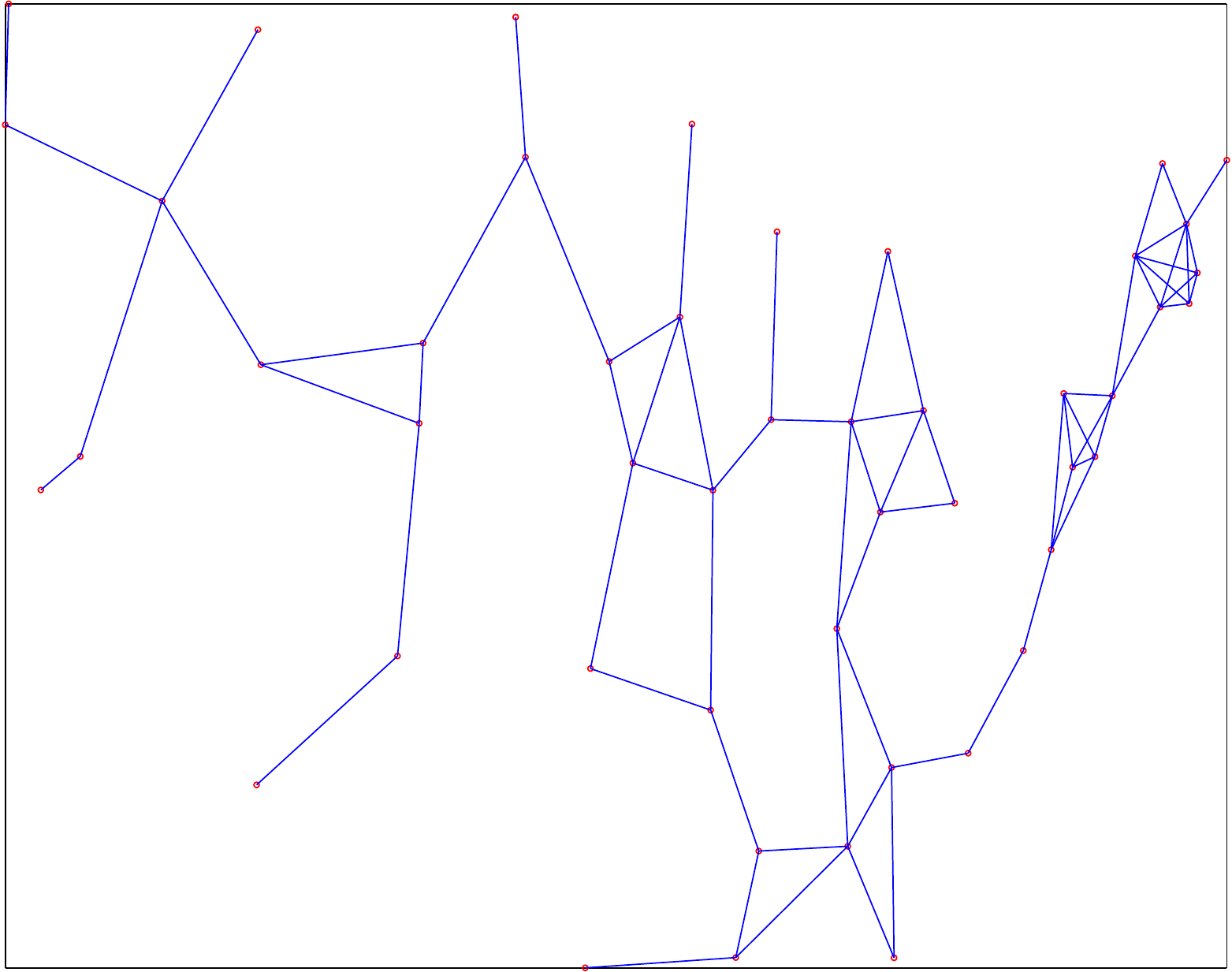}
\label{INUScapitalsxymu3b50OUTpicture}}
\caption{Pictures of connected ESNM on US the state capitals'' locations for two
values of the mean degree.}
\end{figure}

\begin{table}[h]
\centering
\begin{tabular}{|r|c|c|c|c|}
\hline
 & $\mu=2$ & $\mu=2.2$& $\mu=3$ & $\mu=4$\\\hline
mean edge length $\xi$	&   2.70 & 2.66	&2.75  & 2.932  \\
clustering $C$ 		& 0.045 & 0.178 	& 0.425& 0.539\\
hop distance $h $ 	& 9.20 & 8.90	& 6.94 & 5.54  \\
route distance $r$ 	&24.1 & 23.8	& 21.2 & 19.1	\\
Route factor $R$ 	& 0.633	&  0.606 &0.382& 0.234\\\hline
\end{tabular}
\caption{Statistics obtained for the ESNM on US state capitals' locations with
three values of the mean degree and $\beta=50$. \label{capitalstats}}
\end{table}

\begin{figure}[h]
\centering
\subfigure[]{\includegraphics
[width=.47\textwidth]
{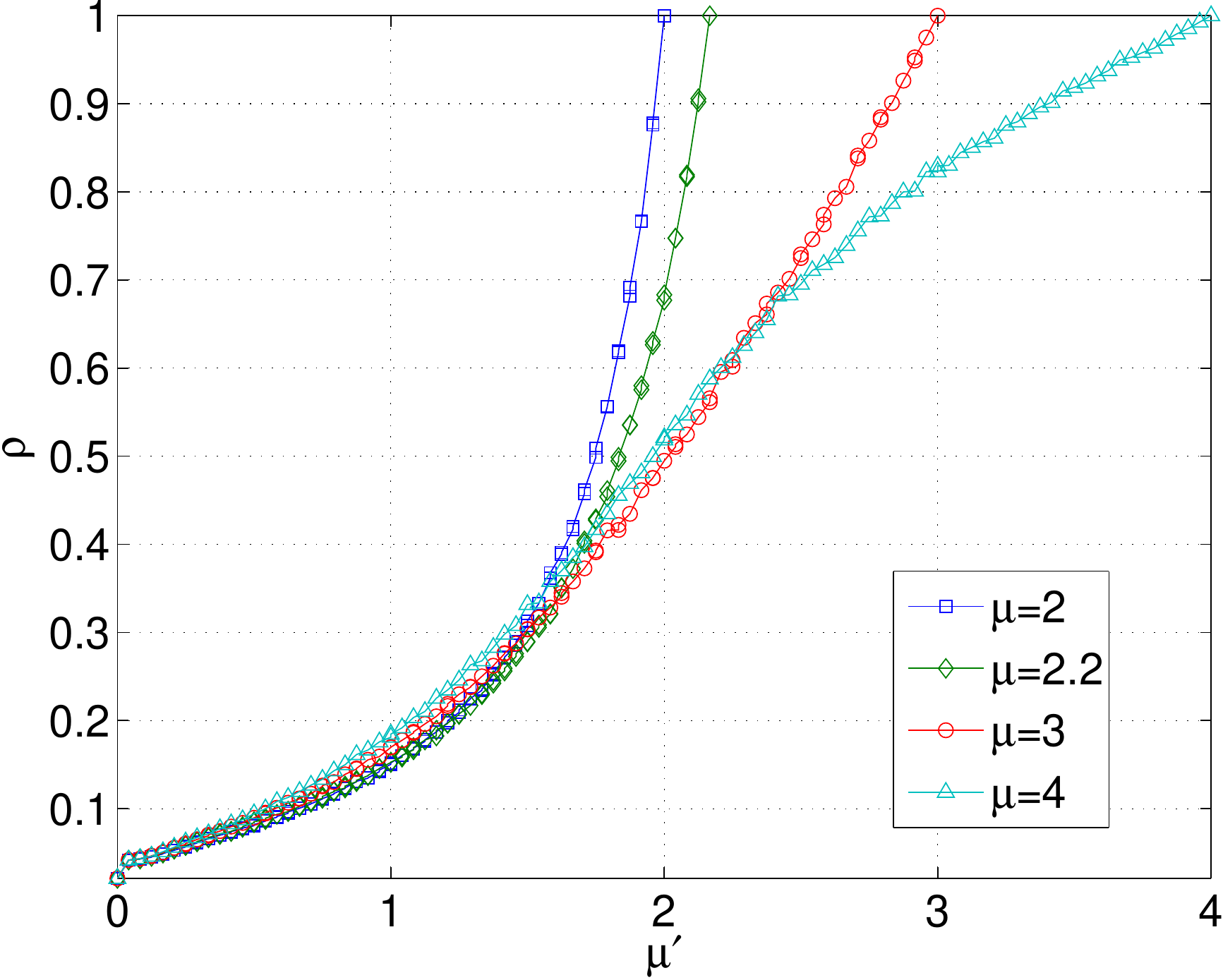} \label{INUScapitalsxymu2mu2_2mu3mu4b50OUTRobustness}}\,
\subfigure[]{\includegraphics
[width=.47\textwidth]
{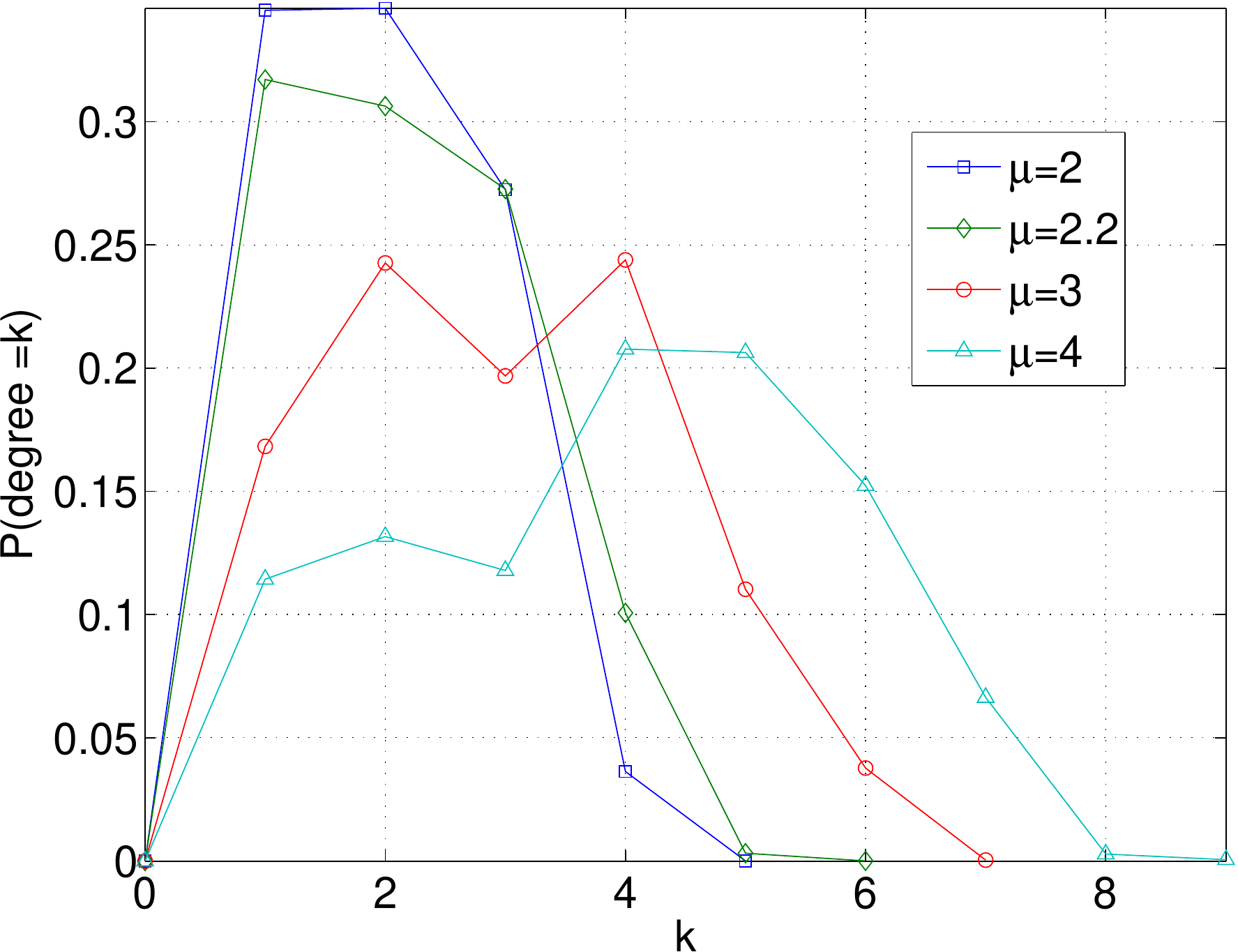} \label{INUScapitalsxymu2mu2_2mu3mu4b50OUTDegreeDist}}
\caption{Comparison of (a) robustness and (b) degree distribution, of simulation
results for US state capitals' locations data for four values of $\mu$.}
\end{figure}

Our next test bed for the connected ESNM is the locations of the lower 48 US
state capitals. Here, the vertices are more uniformly distributed than in the the road
network we considered earlier. 
Nonetheless, \fig{INUScapitalsxymu3b50OUTpicture}{} shows the accumulation of edges
when $\mu=3$, in the north east region where there are many states. 
In Table \ref{capitalstats} shows that, as is to be expected, the clustering
coefficient and all the distance measures decrease when $\mu$ increases. 
The mean edge length, however, is non-monotone, consistent with our earlier
findings shown in \fig{INMon3D2b10k0sOUTu}{}. 
The robustness profile in \fig{INUScapitalsxymu2mu2_2mu3mu4b50OUTRobustness}{} shows
that the $\mu=4$ network is robust by our criterion, while $\mu=2$ and 3 are
not, i.e., $3<\mu_*<4$. 
\fig{INUScapitalsxymu2mu2_2mu3mu4b50OUTDegreeDist}{} shows the degree distribution; 
the bimodality of the curve for $\mu=3$ and 4 is peculiar and is perhaps due to the inhomogeneous distribution of points.
\comment{the
jaggedness of the $\mu =3$ and 4 curves is peculiar and is perhaps due to the
small system size.}

\section{Summary and Outlook}
In this paper we have introduced an abstract model for the evolution of spatial
networks  whose equilibrium distribution is given by an explicit formula containing their spatial dimension, their mean degree, the topological constraint and the inverse temperature $\beta$. 
We examined two cases -- one where the topological constraint is absent, and the other where the network is required to be connected.

The unconstrained network is closely related to a percolation problem. 
This enabled us to analytically compute the 
distribution of the degrees and edge lengths, and the clustering coefficient of the network.
Other quantities such as the critical mean degree $\mu_*^{(\beta \to \infty)}$ for percolation  were estimated by simulation. 
One interesting aspect of this model is that that it interpolates between the \ER random graph ($\beta=0$) and the random geometric graph ($\beta=\infty$). 
Furthermore, the unconstrained ESNM can be a model for a social network where stubborn individuals with fixed 
opinions a number of issues, have a tendency to rewire their ties to those with similar opinions. 
Even when this tendency was high, the fact that the number of edges is fixed, ensured that a small mean degree was enough to 
have a giant component.

An analytical framework for computing quantities associated with the connected network model is lacking, 
so we studied that model purely by simulation concentrating on the random connected network ($\beta=0$) and 
the almost optimized network ($\beta=10$). Our analysis focused on the total length (‘wiring cost’) of the network, 
how routes between vertices compare with their spatial separation, and the robustness of the network to random removal of edges. 
In the former two aspects, we found that the almost optimized network is notably more efficient than the random connected network. 
However, in terms of the metric that we proposed, the RCN was found to be more robust. A peculiar feature we noted of the AON($\mu$) is that the 
mean edge length is the lowest when $\mu \approx 2.2$, and not at 2 as one would expect. 

To test the success of our second model for network design we considered two examples: the Minnesota road network and 
the 48 state capitals of the continental US. While the rewiring produced networks with good values of some important statistics, 
additional criteria (e.g., reweighted edges based on population density to compensate for uneven vertex distributions) will need to be introduced to produce good solutions. 
\comment{
In this paper, we have introduced an abstract model for the evolution of spatial
networks, and investigated some properties of the 
equilibrium network. 
While any realistic study of real networks would require a large number of
parameters, our aim here was to study the simplest model that incorporates
spatial effects. Our model 
therefore has only four parameters. On appealing aspect of the model, we
believe, is its ability to completely ignore space by setting $\beta=0$, leading
to well studied 
networks such as the Erd\H{o}s-R\'{e}nyi random graph and the minimum spanning
tree.
We have also briefly discussed how two networks from this model could be applied
to social and transportation networks 
respectively. 

While for the unconstrained network there exists an equivalent percolation
model, an analytical framework for the study of the connected 
network is lacking and needs to be developed. This could help us better
understand efficient transportation networks. One approach could be to study the network Laplacian
\cite{Samukhin2008} since it encodes the 
connectivity properties of the network (the second smallest eigenvalue is
positive iff the network is connected).}

\begin{appendix}
\section{The grand partition function} \label{chlorine}
It is well known (see for e.g. \cite{Pathria2007}) that the grand partition function of non-interacting particles can be written as a product over single particle states. So, in our case, we have
\begin{align}
 \Xi(\beta,\kappa) = \prod_{\{x,y\} \in \binom{V}{2}} \left( 1 + \kappa e^{-\beta
|x-y|}\right) \,.
\end{align}
Since the partition functions above are conditional on the location of the vertices, we calculate the expected value 
\begin{align}
 \mathbb{E} \log \Xi 
&= \left[\prod_{x\in V} \int_{\mathcal{V}_{nD}}\frac{\mathrm{d}x}{n} \right]\sum_{\{x,y\} \in \binom{V}{2}} \log\left( 1 + \kappa e^{-\beta|x-y|}\right)\nonumber\\
&= \binom{n}{2} \int_{\mathcal{V}_{nD}} \int_{\mathcal{V}_{nD}} \frac{\mathrm{d}x}{n} \frac{\mathrm{d}y}{n} \log\left( 1 + \kappa e^{-\beta|x-y|}\right) \,.\label{ukraine}
\end{align}
In the limit of large $n$, the double integration in \eqref{ukraine} is to be performed over $\mathbb{R}^{2D}$
and all points are identical. We can choose a point at $x$. 
Then calculate $\int g(|x-y|) \mathrm{d}y $, by constructing shells centered at
$x$ at all radii $\varepsilon$. This quantity will be independent of $x$. The remaining 
integral $\int \mathrm{d} x/n $ is just equal to 1, giving
\begin{align}
 \mathbb{E} \log \Xi 
&= \frac{n}{2}  S_{D-1} \int_{0}^\infty  \varepsilon^{D-1} \log\left( 1 + \kappa e^{-\beta \varepsilon}\right)    \mathrm{d}\varepsilon  = \frac{n}{2} \frac{S_{D-1} \Gamma(D)}{\beta^D} [-\mathrm{Li}_{D+1} (-\kappa)]\,.\label{ukraine2}
\end{align}

\section{Degree distribution in the percolation network} \label{sodium}
We can calculate the degree distribution of the percolation network as follows.
Let $X$ be a randomly chosen vertex and let $Y$ be one of the other vertices. The probability that $X$ is connected to $Y$ is 
\begin{align*}
\mathbb{P}(\{X,Y\} \in E ) = \int \mathbb{P}(\{X,Y\} \in E | Y=y) \mathbb{P}(Y=y) = \frac{1}{n} \int g(|X-y|)\mathrm{d}y\,.
\end{align*}
The probability that $X$ is connected to exactly $k$ of the other $n-1$ vertices is 
\begin{align}
\mathbb{P}(d(X)=k) &= \mathrm{Binomial}\left(n-1,\mathbb{P}(\{X,Y\} \in E );k \right) \nonumber\\
&= \mathrm{Binomial}\left(n-1,\frac{1}{n} \int g(|X-y|)\mathrm{d}y;k \right) \nonumber\\
&\to \mathrm{Poisson}\left(\int g(|X-y|)\mathrm{d}y;k \right)\quad \textrm{as} \quad n\to \infty\,.
 \end{align}
So a vertex at $x$ has a degree distribution $\mathrm{Poisson}[\mu(x)]$ where $\mu(x) = \int g(|x-y|)\mathrm{d}y$.

\section{Clustering Coefficient for $\beta \to \infty$} \label{potassium}
Consider a vertex $z$ that is connected to vertices $x$ and $y$.  This means that $x$ and $y$ lie within a $D-$ball of radius $\varepsilon_{0}$ centered at $z$. 
Now, $x$ and $y$ will be connected to each other only if $y$ lies in the intersection of the $D-$balls of radii $\varepsilon_{0}$ centered at $x$ and $z$ respectively. In other words, the 
probability that $y$ is connected to $x$ is the ratio of the intersection volume of the $D-$balls to the volume of a $D-$ball. The volume of the cap that subtends a half angle $\theta$ of a unit $D$-ball is
given by  
\begin{align}
 \Omega_D^{\textrm{cap}}(\theta) = \frac{\pi^{\frac{D-1}{2}}}{\Gamma\left(
\frac{D+1}{2}\right)} \int_0^{\theta} \sin^D t \,\,\mathrm{d} t\,.
\end{align}
If $|x-z| =\varepsilon <\varepsilon_{0}$, to find the intersection volume, we need to add the volumes of two such caps
with $\theta = \arccos(\varepsilon/2\varepsilon_{0})$. The probability that $|x-z| =\varepsilon $ is $S_{D-1} \varepsilon^{D-1} \mathrm{d}\varepsilon/\Omega_D\varepsilon_{0}^D$. So averaging the
intersection volume over all $\varepsilon$, we have,
\begin{align}
 C &= \frac{1}{\Omega_D \varepsilon_{0}^D}\int_0^{\varepsilon_{0}} 2 \Omega_D^{\textrm{cap}}\left(\arccos\left(\frac{\varepsilon}{2\varepsilon_{0}}\right)\right) \varepsilon_{0}^D
\frac{S_{D-1} \varepsilon^{D-1} \mathrm{d}\varepsilon}{\Omega_D\varepsilon_{0}^D}\nonumber\\
&= \frac{2D^2}{\sqrt{\pi}} \frac{\Gamma(D/2)}{\Gamma((D+1)/2)} \int_0^1  \int_0^{\arccos(t/2)} \sin^D \tau \,\mathrm{d}\tau \, t^{D-1}  \,\mathrm{d}t\,.
\end{align}
\comment{
\section{Hop and route distances and the route factor} \label{natrium}
Consider the set of paths $\mathcal{P}(x,y) = \{ (x,i_1,i_2,\ldots,i_{-1},y):\{x,i_1\}, \{i_1,i_2\},\ldots,\{i_{-1},y\} \in E\}$ between two vertices $x$ and $y$. 
For a path $P \in \mathcal{P}(x,y) $, the hop length $h(P)$ is the number of edges in $P$, and 
the route length $r(P) = |x-i_{1}| + |i_{1}
-i_{2}|+\ldots+|i_{-1}-y|$. Related to these path lengths, are two kinds of distances between $x$ and $y$: the hop distance $h(x,y) = \min_{P\in \mathcal{P}(x,y)} h(P) $, 
and the route distance $r(x,y) = \min_{P\in \mathcal{P}(x,y)} r(P)$. The
route factor \cite{Aldous2010} is defined as 
\begin{align}
 R(x,y) = \frac{r(x,y)}{|x-y|} -1\,.
\end{align}
We study these quantities averaged over all vertex
pairs of the network.
}
\end{appendix}

\bibliographystyle{unsrt}	
\bibliography{/home/varghese/Dissertation/Bibliography/mybib}
\end{document}